\documentclass[10pt]{article}
\usepackage[margin=1.2in]{geometry}
\usepackage{hyperref}
\usepackage{graphicx}
\usepackage{dsfont}
\usepackage{enumitem}
\usepackage{tikz}
\usepackage{caption}
\usetikzlibrary{automata,arrows,positioning,calc}

\usepackage[thinlines]{easytable} 
\usepackage{amsthm} 
\usepackage{amsmath}
\usepackage{amssymb}
\usepackage{mathtools}

\newtheorem{theorem}{Theorem}
\newtheorem{remark}[theorem]{Remark}
\newtheorem{definition}[theorem]{Definition}

\newtheorem{lemma}[theorem]{Lemma}

\newtheorem{example}[theorem]{Example}
\newtheorem{prop}[theorem]{Proposition}

\newcommand{\R}{\mathbb{R}}
\newcommand{\N}{\mathbb{N}}

\newcommand{\veps}{\varepsilon}

\newcommand{\Nodd}{\N^{\text{odd}}}
\newcommand{\Neven}{\N^{\text{even}}}

\newcommand{\beq}{\begin{equation*}}
\newcommand{\eeq}{\end{equation*}}
\newcommand{\head}[1]
{\markright{\hbox to 0pt{\vtop to 0pt{\hbox{}\vskip 3mm \hrule width
\textwidth \vss} \hss}{\sc #1}}}

\begin{document}


\parindent=0pt
\baselineskip18pt
\parskip6pt
\head{Search problem}

\title{A competitive search game with a moving target\thanks{We would like to thank Steve Alpern, Miquel Oliu-Barton and J\'er\^ome Renault for their precious comments and referring us to the related literature.} \thanks{Declarations
of interest: none}}
\date{\today}
\author{Benoit Duvocelle  \footnote{Corresponding author. Address: School of Business and Economics, Quantitative Department, Tongersestraat 53, 6211 LM, Maastricht, The Netherlands} \thanks{Department of Quantitative Economics, Maastricht University. Email: benoit.duvocelle@hotmail.fr}\hskip6pt
J\'anos Flesch\thanks{Department of Quantitative Economics, Maastricht University. Email: j.flesch@maastrichtuniversity.nl} \hskip6pt
Mathias Staudigl\thanks{Department of Data Science and Knowledge Engineering, Maastricht University. Email: m.staudigl@maastrichtuniversity.nl} \hskip6pt
Dries Vermeulen\,  \thanks{Department of Quantitative Economics, Maastricht University. Email: d.vermeulen@maastrichtuniversity.nl}  \hskip6pt}  
\maketitle

\begin{abstract}
We introduce a discrete-time search game, in which two players compete to find an object first. The object moves according to a time-varying Markov chain on finitely many states. The players know the Markov chain and the initial probability distribution of the object, but do not observe the current state of the object. The players are active in turns. The active player chooses a state, and this choice is observed by the other player. If the object is in the chosen state, this player wins and the game ends. Otherwise, the object moves according to the Markov chain and the game continues at the next period. 

We show that this game admits a value, and for any error-term $\veps>0$, each player has a pure (subgame-perfect) $\veps$-optimal strategy. Interestingly, a 0-optimal strategy does not always exist. The $\veps$-optimal strategies are robust in the sense that they are $2\veps$-optimal on all finite but sufficiently long horizons, and also $2\veps$-optimal in the discounted version of the game provided that the discount factor is close to 1. We derive results on the analytic and structural properties of the value and the $\veps$-optimal strategies. Moreover, we examine the performance of the finite truncation strategies, which are easy to calculate and to implement. We devote special attention to the important time-homogeneous case, where additional results hold.
\end{abstract}

\textbf{Keywords:} Search game; sequential game; two-player zero-sum game; subgame perfect $\veps$-equilibrium; discrete time-varying Markov process.

\section{Introduction}
The field of search problems is one of the original disciplines of Operations Research. 
In the basic settings, the searcher's goal is to find a hidden object, also called the target either, with maximal probability or as soon as possible. 
By now, the field of search problems has produced into a wide range of models. 
The models in the literature differ from each other by the characteristics of the searchers and of the objects. 
Concerning objects, there might be one or several objects, mobile or not, and they might have no aim or their aim is to not be found. 
Concerning the searchers, there might be one or more. 
When there is only one searcher, the searcher faces an optimization problem. 
When there are more than one searcher, searchers might be cooperative or not.
If the searchers cooperate, their aim is similar to the settings with one player: they might want to minimize the expected time of search, the worst time, or some search cost function. 
If the searchers do not cooperate, the problem becomes a search game with at least two strategic non-cooperative players, and hence game theoretic solution concepts and arguments will play an important role. For an introduction to search games, we refer to \cite{alpern2006theory}, \cite{gal1979search}, \cite{gal2010search}, \cite{gal2013search}, \cite{garnaev2012search},  and for surveys see \cite{benkoski1991survey} and  \cite{hohzaki2016search}.

We introduce a competitive search game, played at discrete periods in $\N$. An object is moving according to a time-varying Markov chain on finitely many states. 
Two players compete to find the object first. 
Consider for example two pharmaceutical firms which compete in developing a cure for a disease that mutates over time.
They both know the Markov chain and the initial probability distribution of the object, but do not observe the current state of the object. Player 1 is active at odd periods, and player 2 is active at even periods. The active player chooses a state, and this choice is observed by the other player. If the object is in the chosen state, this player wins and the game ends. Otherwise, the object moves according to the Markov chain and the game continues at the next period. If the object is never found, the game lasts indefinitely. In that case, neither player wins.

When the active player chooses a state, he needs to take two opposing effects into account. First, if the object is at the chosen state, then he wins immediately. This aspect makes choosing states favorable where the object is located with a high probability. Second, if the object is not at the chosen state, then knowing this, the opponent gains information: the opponent can calculate the conditional probability distribution of the location of the object at the next period. This aspect makes choosing states favorable where, on condition that the object not being there, the induced conditional distribution at the next period disfavors the opponent. In particular, this conditional distribution should not be too informative, and for example it should not place too high a probability on a state. Clearly, in some cases there is no state that would be optimal for both scenarios at the same time, and hence the active player somehow needs to aggregate the two scenarios in order to make a choice.

Each player's goal is to maximize the probability to win the game, that is, to find the object first. In our model, we do not assume that the players take into account the period when the object is found. Of course, in most cases, maximizing the probability to win will entail at least partially that each player would prefer to find the object at earlier periods, thereby preventing the other player from finding the object. We refer to Section \ref{additionalresults} on the finite horizon and on the discounted versions of the search game, where the period when the object is found also matters.

The two players have opposite interests, up to the event when the object is never found. More precisely, each player's preferred outcome is that he finds the object, but he is indifferent between the outcome that the other player finds the object and the outcome that the object is never found. As we will see, the possibility that neither player finds the object will only have minor role, and hence the two players have essentially opposite interests in the search game.

\textbf{Main results.} Our main results can be summarized as follows.

[1] We study the existence of $\veps$-equilibria. A strategy profile is called an $\veps$-equilibrium if neither player can increase his expected payoff by more than $\veps$ with a unilateral deviation. We prove that each competitive search game admits an $\veps$-equilibrium in pure strategies, for all error-terms $\veps>0$ (cf. Theorem~\ref{vepsequilibrium} and for subgame-perfect $\veps$-strategies cf. Proposition \ref{subgameperfectvepsequilibrium}). The proof is based on topological properties of the game. Interestingly, a 0-equilibrium does not always exist, not even in mixed strategies. We demonstrate it with two different examples (cf. Examples~\ref{noeq1} and~\ref{noeq2}). 

[2] We examine the properties of $\veps$-equilibria. We show that in each $\veps$-equilibrium, the object is eventually found with probability at least $1-\veps\cdot |S|$, where $|S|$ is the number of states (cf. Lemma~\ref{bestresponse}), and that the set of $\veps$-equilibrium payoffs converge to a singleton $(v,1-v)$, with $v\in[\tfrac{1}{|S|},1]$ as $\veps$ vanishes (cf. Proposition~\ref{valueexist} and Theorem~\ref{vepsnashvalue}). This implies that, in such search games, the two players have essentially opposite interests, and that we may consider $v$ to be the value of the game and the strategies of $\veps$-equilibria as $\veps$-optimal strategies (cf. Definition~\ref{vepsoptimaldef} and Proposition~\ref{vepsoptimalprop}). 

[3] We prove that the $\veps$-optimal strategies are robust in the following sense: they are $2\veps$-optimal if the horizon of the game is finite but sufficiently long (cf. Theorem~\ref{theorem-fin-hor}), and they are also $2\veps$-optimal in the discounted version of the game, provided that the discount factor is close to 1 (cf. Theorem~\ref{discount}).

[4] We investigate the functional and structural properties of the value and the $\veps$-optimal strategies (cf. Theorems~\ref{linear},~\ref{lip} and~\ref{geometry}). In particular, we consider the set of probability distributions for the location of the object where choosing a particular state is optimal, and show that this set is star-shaped.

[5] Since the $\veps$-optimal strategies may have a complex structure and may be difficult to identify, we examine the finite truncation strategies, which maximize the probability to win in a finite number of periods. We show that each finite truncation strategy, provided that the horizon of the truncation is sufficiently long, is $\veps$-optimal in the search game on the infinite horizon (cf Theorem~\ref{theorem-fin-hor}). Note that the finite truncation strategies are easy to calculate by backward induction and only require finite memory.  

[6] We devote attention to the special case when the Markov chain is time-homogenous (cf. Section~\ref{timehomogeneous}), as time-homogenous Markov chains are well studied in the literature of Markov chains and frequently used in applications. For time-homogenous Markov chains, we prove additional results. In particular, if the initial probability distribution of the object is an invariant distribution of the time-homogenous Markov chain, then the value is at least 1/2, so player 1 has a weak advantage (cf. Proposition~\ref{invariantdistribution}). Moreover, if the time-homogenous Markov chain is irreducible and aperiodic, then the game admits a 0-equilibrium in pure strategies (cf. Theorem~\ref{ergodic}). \medskip

\textbf{Related literature}

Discrete search problems with a moving object have been widely investigated. \cite{pollock1970simple}, \cite{schweitzer1971threshold}, \cite{dobbie1974two} and  \cite{kan1974counterexample} study the two-state problem. Assuming perfect detection, \cite{nakai1973model} investigates the three-state problem. \cite{brown1980optimal} considers the search for a target with Markov motion in discrete time and space using an exponential detection function. 
He provides a necessary and sufficient condition for an optimal search plan and an efficient iterative algorithm for generating optimal plans. \cite{washburn1983search} studies a discrete effort analogue of \cite{brown1980optimal}, in which searchers decide the effort they want to invest in order to find the object at each location they visit. General necessary and sufficient conditions which extend Brown's results to an arbitrary stochastic process for any mixture of discrete and continuous time and space are given in \cite{stone1976theory}. 
More recently, \cite{garrec2020search} study a hide-search game in a random graph, that is a graph in which each edge is available at each period with a positive probability. For extensive surveys, see \cite{benkoski1991survey} and \cite{hohzaki2016search}.

Most of the search games focus on the case of one searcher, or several cooperative searchers. Some problems with several cooperative searchers and one or several moving targets are mentioned in the book of \cite{stone2016optimal}, where some algorithms are also studied to solve those problems. To the best of our knowledge, only two models consider several competitive searchers. \cite{nakai1986search} investigates a non-zero-sum game in which two searchers compete with each other for quicker detection of an object hidden in one of $n$ boxes, with exponential detection functions. Each player wishes to maximize the probability that he detects the object before the opponent detects it. The author shows the existence of an equilibrium point of the form of a solution of simultaneous differential equations, and gets explicit solution results showing that both players have the same equilibrium strategy even though the detection rates are different.
\cite{flesch2009optimal} investigate the problem in which an agent has to find an object that moves
between two locations according to a discrete Markov process, with the additional costless option to wait instead of searching. They find a unique optimal
strategy characterized by two thresholds and show that, in a clear contrast with our model, it can never be optimal to search the location with the lower probability of containing the object. They also analyze the case of multiple agents, where the agents not only compete against time but also against each other in finding the object. They find different kinds of subgame
perfect equilibria.

As in \cite{nakai1973model} we investigate functional and structural properties of the objective function. Nakai proved that the function that allocates to a probability distribution the average number of looks before finding the object is continuous, concave and enjoy some linear properties. They also show that the optimal decision regions (see Section~\ref{geometry}) are star-convex. These properties have also been studied in \cite{macphee1995optimal} and in the PhD thesis of \cite{jordan1997optimal}.

\textbf{Structure of the paper.} In Section~\ref{model}, we present the model. In Section~\ref{existence}, we examine the existence of $\veps$-equilibrium, for $\veps\geq 0$. In Section~\ref{payoffproperties}, we argue that the two players have essentially opposite interests, and we define the value and the notion of $\veps$-optimal strategies. In Section~\ref{strategies}, we present two relevant strategies, namely the finite truncation strategy and the discounted strategy, and we prove payoff guarantees of those strategies. In Section~\ref{additionalresults} we present additional results related to the structural properties of the value, the subgame-perfect equilibria and the case in which the Markov chain is time-homogeneous. Functional properties of the value can be found in the Appendix. The conclusion is in Section~\ref{conclusion}.

\section{The Model}\label{model}

\textbf{The Game.} We study a competitive search game $G$ played by two players. Let $\N=\{1,2,3,\ldots\}$. An object is moving according to a discrete-time Markov chain $(X_t)_{t\in \N }$ on a finite state space $S$. The initial probability distribution of the object over the set $S$ is given by $p\in\Delta(S)$, and the transition probabilities at period $t$ are given by an $S \times S$ transition matrix $P_t=[P_t(i,j)]_{(i,j)\in S^2}$, where $P_t(i,j)$ is the probability for the object to move from state $i$ to state $j$ at period $t$.

At each period $t\in\N$, one of the players is active: At odd periods player 1 is the active player, and at even periods player 2 is the active player. 
The active player chooses a state $s_t\in S$, which we call the action at period $t$. If the object is at state $X_t=s_t$, then the active player finds the object and wins the game. 
Otherwise, the object moves according to the transition matrix $P_t$ at time $t$ and the game enters period $t+1$. 
We assume that each player observes the actions chosen by his opponent and each player is aware of which actions the himself has chosen in the past.
The transition matrices $(P_t)_{t\in \N}$ and the initial distribution $p$ are known to the players.

The aim of each player is to maximize the probability that he finds the object first. 

\textbf{Histories.} A history at period $t\in \N$ is a sequence $h_t=(s_1,\ldots,s_{t-1})\in S^{t-1}$ of past actions. By $H_t=S^{t-1}$ we denote the set of all histories at period $t$. Note that $H_1$ consists of the empty sequence. Let $\N^{\text{odd}}=\{1,3,5,\ldots\}$ and $\N^{\text{even}}=\{2,4,6,\ldots\}$. We denote by $H^{\text{odd}}=\cup_{t\in \N^{\text{odd}}}H_t$ the set of histories at odd periods, and by $H^{\text{even}}=\cup_{t\in \N^{\text{even}}}H_t$ the set of histories at  even periods. Note that at each history $h$, the players can calculate the probability distribution for the current location of the object. 
\vskip6pt

\textbf{Strategies.} The action sets for both players are $A_1=A_2=S$. A strategy $\sigma = (\sigma_t)_{t\in \N^{\text{odd}}}$ for player 1 is a sequence of functions $\sigma_t \colon H_t\rightarrow \Delta(S)$. The interpretation is that, at each period $t\in \N^{\text{odd}}$, given the history $h_t$, the strategy $\sigma_t$ chooses to search state $s\in S$ with probability $\sigma_t(h_t)(s)$. Similarly, a strategy $\tau = (\tau_t)_{t\in \N^{\text{even}}}$ for player 2 is a sequence of functions $\tau_t \colon H_t\rightarrow \Delta(S)$. We denote by $\Sigma$ and $\mathcal{T}$ the set of strategies for players 1 and 2, respectively. Note that $\Sigma=\prod_{h\in H^{\text{odd}}}\Delta(S)$ and $\mathcal{T}=\prod_{h\in H^{\text{even}}}\Delta(S)$. We say that a strategy is pure if, for any history, it places probability 1 on one action.

\vskip6pt

\textbf{Winning probabilities.}
We define the stopping time\footnote{With the convention that $\min\{\emptyset\}=+\infty$} of the game by  $\Theta=\min\{t \in \N| \ s_t=X_t\}$. Consider a strategy profile $(\sigma,\tau)$. The probability under $(\sigma,\tau)$ that player 1 wins is denoted by $u_1(\sigma,\tau)=\mathbb{P}_{\sigma,\tau}\left(\Theta\in \N^{\text{odd}}\right)$, and that player 2 wins is denoted by $u_2(\sigma,\tau)=\mathbb{P}_{\sigma,\tau}\left(\Theta\in \N^{\text{even}}\right)$. Note that $u_1(\sigma,\tau)+u_2(\sigma,\tau)=1-\mathbb{P}_{\sigma,\tau}(\Theta=\infty)$. If the object has not been found before period $t$, and the history is $h_t$, the continuation winning probabilities from period $t$ onward are denoted by $u_1(\sigma,\tau)(h_t)$ for player 1 and $u_2(\sigma,\tau)(h_t)$ for player 2. \footnote{When we wish to emphasize the parameter $p$, we will write $u_1(\sigma,\tau)(p)$ and $u_2(\sigma,\tau)(p)$.}

\textbf{$\veps$-Equilibrium.} Let $\veps\geq 0$ be an error-term. A strategy $\sigma$ for player 1 is an $\veps$-best response against strategy $\tau$ for player 2 if $u_1(\sigma,\tau) \geq u_1(\sigma',\tau)-\veps$ for every strategy $\sigma'$ of player 1. Similarly, a strategy $\tau$ for player 2 is an $\veps$-best response against strategy $\sigma$ for player 1 if $u_2(\sigma,\tau) \geq u_2(\sigma,\tau')-\veps$ for every strategy $\tau'$ of player 2. A strategy profile $(\sigma,\tau)$ is called an $\veps$-equilibrium if $\sigma$ is an $\veps$-best response against $\tau$ and $\tau$ is an $\veps$-best response against $\sigma$.

\textbf{An alternative interpretation of the game.} We call the previous game Model [1]. We present an alternative model of this game in perfect information. This model is useful in order to prove the existence of $\veps$-equilibrium for all $\veps>0$ (cf. Theorem \ref{vepsequilibrium}).

[2] Another way to describe our game is as follows. One could imagine that the game consists of two phases. In the first phase the players choose actions. More precisely, in the first phase player 1 chooses an action at odd periods and player 2 chooses an action at even periods just as before. This results in an infinite sequence of states $(s_1,s_2,\ldots)$. The set of infinite histories is $S^\infty$. Every pure strategy profile $(\sigma,\tau)$ induces a unique infinite history $h^\infty_{\sigma,\tau}\in S^\infty$. In a second phase, players receive a payoff. Now, for $i=1,2$, consider the payoff function $f_i:S^\infty\to [0,1]$ defined as follows. Consider an infinite history $(s_1,s_2,\ldots)$. Take any pure strategy profile $(\sigma,\tau)$ such that $h^\infty_{\sigma,\tau}=(s_1,s_2,\ldots)$ and define $f_i(s_1,s_2,\ldots)=u_i(\sigma,\tau)$. 
Note that this definition only depends on the realized history.
The goal of each player is to maximize his payoff. Note that this is a game without an object. This way we obtain a two-player perfect information game.

\textbf{Discussion.} We briefly argue that the above descriptions are equivalent. For each pure strategy profile $(\sigma,\tau)$, for each player $i=1,2$, we have $u_i(\sigma,\tau)=f_i(h^\infty_{\sigma,\tau})$. Then, a strategy profile in one of the models leads to the same payoff in the other game. The difference is that Model [1] is in imperfect information, as players only know the probability distribution of the object, while Model [2] is in perfect information.

Model [1] gives a very clear, intuitive and concrete description of the game. This is the reason why we usually work with this model in the paper. Model [2] is used as a tool to prove existence of $\veps$-equilibrium as in Theorem \ref{vepsequilibrium}. 

\section{Existence of equilibrium} \label{existence}
In this section, we examine equilibria in competitive search games. In the first subsection, we show that there are search games for which there exist no 0-equilibrium, not even in mixed strategies. 
From a technical point of view, this is caused by discontinuity in the payoff functions of the players.
In the second subsection, we focus on the notion of $\veps$-equilibrium, where $\veps>0$ is an error-term, and prove that each search game admits an $\veps$-equilibrium in pure strategies, for all $\veps>0$. 
We conclude the section by presenting an  $\veps$-equilibrium for the games introduced in the first subsection.

\subsection{Search games with no 0-equilibrium}

\begin{theorem}\label{noeqthm} There exist time-homogeneous competitive search games which admit no 0-equilibrium, not even in mixed strategies.
\end{theorem}

We provide two counter-examples: Example~\ref{noeq1} and Example~\ref{noeq2}. A common property of these counter-examples is that during the game the players are forced to choose states where the probability of the object is positive but converges to zero when $t$ goes to infinity. In Example~\ref{noeq1}, this happens within the class of transient states. In contrast, in Example~\ref{noeq2}, there are multiple ergodic sets in the Markov chain, and the players have an incentive to choose states in an ergodic set, even when the conditional probability that the object is in this ergodic set is very small.

\begin{example} \label{noeq1}\rm Consider the game in Figure 1. In this game, $\eta\in (0,1/4)$ and the initial probability distribution is $p=(q,q,1/2-q,1/2-q)$, where $q\in (0,1/4)$. Notice that states 1 and 2 have the same transition probabilities, and so do states 3 and 4. States 1 and 2 are transient, whereas states 3 and 4 are absorbing.

\begin{figure}[htbp] \label{noeq1figure}
	\centering
\begin{tikzpicture}[->, >=stealth', auto, semithick, node distance=3cm]
\tikzstyle{every state}=[fill=white,draw=black,thick,text=black,scale=1]
\node[state]    (1)               {$1$};
\node[state]    (2)[right of=1]   {$2$};
\node[state]    (3)[below of=1]   {$3$};
\node[state]    (4)[below of=2]   {$4$};
\path
(1) edge[loop left]   node{$\frac{\eta}{2}$}     (1)
    edge[bend left]   node{$\frac{\eta}{2}$}     (2)
    edge[bend right]  node{$\frac{1-\eta}{2}$}     (3)
    edge[bend right]              node{$\frac{1-\eta}{2}$}     (4)		
(2) edge[loop right]   node{$\frac{\eta}{2}$}     (2)
    edge[bend left]   node{$\frac{\eta}{2}$}     (1)
    edge[bend left]              node{$\frac{1-\eta}{2}$}     (3)
    edge[bend left]  node{$\frac{1-\eta}{2}$}     (4)	
(3) edge[loop below]  node{$1$}     (3)
(4) edge[loop below]  node{$1$}     (4);
\end{tikzpicture}
  \caption{A game without 0-equilibrium.}
\end{figure}
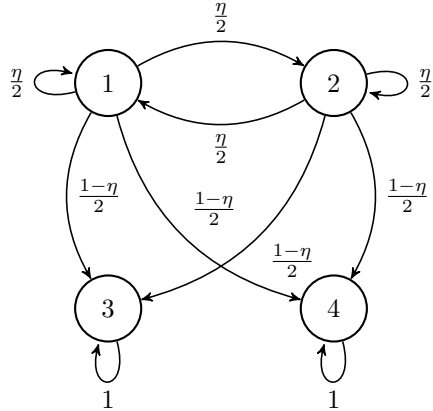

We show that this game admits no 0-equilibrium. The intuition for this claim is as follows. As we will show, it is not optimal for either player to be the first one who chooses an absorbing state. As a consequence, both players prefer to choose state 1 or state 2 and wait until the other player chooses state 3 or state 4. However, if both players do so, they will choose state 1 and state 2 forever, which is not a 0-equilibrium.

Let $\sigma=(\sigma_t)_{t\in\Nodd}$ be the strategy of player 1 defined as follows. For all $t\in \Nodd$, for all $h_t\in H_t$,

\begin{align*}
\sigma_t(h_t)=\left\{ \begin{array}{lll} 
\mbox{state } 1 &&\mbox{ if } t=1, \\
\mbox{state } 3 &&\mbox{ if } t\geq 3 \mbox{ and } h_t(t-1)= 4 \\
&\mbox{ or } &\mbox{ if } t\geq 3 \mbox{ and } h_t(t-1)\in \{1,2\} \mbox{ and } h_t(t-2)\neq 3, \\
\mbox{state } 4 &&\mbox{ if } t\geq 3 \mbox{ and } h_t(t-1)= 3 \\
&\mbox{ or } &\mbox{ if } t\geq 3 \mbox{ and } h_t(t-1)\in \{1,2\} \mbox{ and } h_t(t-2)=3, \\
\end{array}
\right.
\end{align*}
where $h_t(t-2)$ and $h_t(t-1)$ are the second-to-last and the last actions chosen under history $h_t$, respectively.
The idea is that from period 3 onward, $\sigma$ chooses the most likely location of the object.

\textsc{Claim 1:} When player 1 uses $\sigma$ he guarantees himself strictly more than 1/2: $u_1(\sigma,\tau)>1/2$ for every $\tau$.

\textsc{Proof of Claim 1:} 
Under $\sigma$, player 1 looks at state 1 at period 1 and finds the object with probability $q$ at period 1.
If the object is not found, there is a positive probability that it is in state 2 at period 1, in which case it moves with positive probability to state 3 or state 4 at period 2. 
Then player 1 looks at state 3 or state 4 at period 3, depending on the action of player 2 at period 2, and finds the object with probability strictly greater than $1/2-q$ at period 3 no matter the action of player 2 at period 2. \qed

\textsc{Claim 2:} Suppose that player 1 follows a strategy $\sigma$ that looks at state 3 or state 4 at period 1. Then player 2 has a strategy $\tau$ such that $u_1(\sigma,\tau)\leq 1/2$. 

\textsc{Proof of Claim 2:} Let $\tau=(\tau_t)_{t\in\Neven}$ be the strategy of player 2 defined as follows. For all $t\in \Neven$, for all $h_t\in H_t$,
\begin{align*}
\tau_t(h_t)=\left\{ \begin{array}{lll} 
                        \mbox{state } 1 &&\mbox{ if } h_t\in \{1,2\}^{t-1}, \\
                        \mbox{state } 3 &&\mbox{ if } h_t(t-1)= 4 \\
                        &\mbox{ or } &\mbox{ if } h_t(t-1)\in \{1,2\} \mbox{ and } h_t(t-2)=4, \\
					    \mbox{state } 4 &&\mbox{ if } h_t(t-1)= 3 \\
					    &\mbox{ or }&\mbox{ if } h_t(t-1)\in \{1,2\} \mbox{ and } h_t(t-2)=3. \\
					\end{array}
			\right.
\end{align*}
The idea is that $\tau$ looks at state 1 if player 1 has never played state 3 or state 4, and plays the most likely state otherwise.
Assume for simplicity that player 1 looks at state 3 at period 1. Assume that player 1 does not find the object at period 1. 
The conditional probability for the object of being in state 4 at period 2 is then equal to 
\[ p_2(4)=\frac{1/2-q}{1/2+q}+2\cdot \frac{1-\eta}{2}\cdot \frac{q}{1/2+q}=\frac{1/2-\eta\cdot q}{1/2+q},\]
which is strictly higher than 1/2 by our assumption that $q<1/4$ and $\eta<1/4$. 
Then, in the continuation of the game, player 2 guarantees strictly more than 1/2 if he looks at state 4 at period 2. 
If he does not, player 2 will get strictly less than 1/2 if player 1 looks at state 4 at period 3. 
For similar reasons, if period 3 is reached, it is better for player 1 to look at state 3. By repeating this argument, it is better for player 1 to always look at state 3 against $\tau$.

At period 1, player 1 finds the object with probability $1/2-q$. 
At period 2, player 2 finds the object with probability $1/2-q+q(1-\eta)$.
At period 3, player 1 finds the object with probability $q(1-\eta)+q(1-\eta)\eta$. 
At period 4, player 2 finds the object with probability $q(1-\eta)\eta+q(1-\eta)\eta^2$. And so on. Then, player 1 finds the object with probability
\[\frac{1}{2}-q+q(1-\eta)+q(1-\eta)\eta+q(1-\eta)\eta^2+q(1-\eta)\eta^3+\ldots=\frac{1}{2}-q+q(1-\eta)\frac{1}{1-\eta}=\frac{1}{2}.\] 
So, by playing state 3 or state 4 at period 1, player 1 gets at most $1/2$ against $\tau$. \qed

\textsc{Claim 3:} There is no 0-equilibrium. 

\textsc{Proof of Claim 3:} Assume by way of contradiction that there is a 0-equilibrium $(\sigma',\tau')$. From \textsc{Claim 1} and \textsc{Claim 2}, player 1 chooses state 1 or state 2 with probability 1 at period 1. In both cases, at period 2 the current probability distribution is $p_2=(q\eta,q\eta,1/2-q\eta,1/2-q\eta)$. Then, at period 2, the game is similar to the original one, with a parameter $q'=q\eta$ instead of $q$, which still satisfies $q'\in (0,1/4)$, and where the roles of the players are exchanged. Then, as $\tau$ is a 0-best response, it follows from the previous reasoning that player 2 plays state 1 or state 2 with probability 1. By following this process recursively, players will choose states 1 and 2 with probability 1 forever. This leads to the payoff $\frac{4}{4-\eta^2}$ for player 1. Then, player 1 has an incentive to deviate from $\sigma'$ and to choose state 3 at period 1 to get a payoff of at least $1/2-q>\frac{4}{4-\eta^2}$, a contradiction. \qed
\end{example}

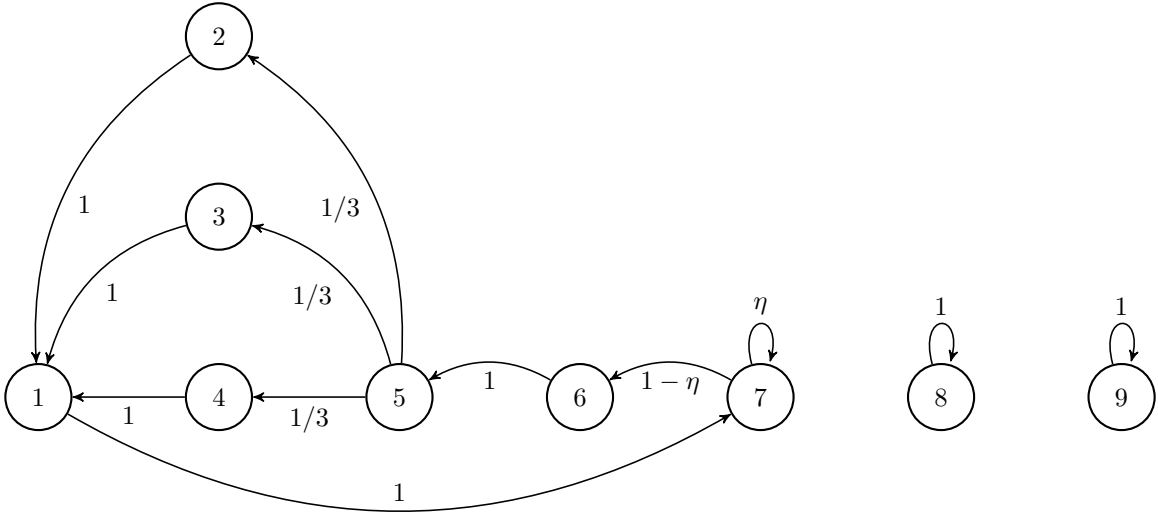
\begin{figure}[htbp] 
	\centering
\begin{tikzpicture}[->, >=stealth', auto, semithick, node distance=2.4cm]
\tikzstyle{every state}=[fill=white,draw=black,thick,text=black,scale=1]
\node[state]    (1)               {$1$};
\node[state]    (4)[right of=1]   {$4$};
\node[state]    (3)[above of=4]   {$3$};
\node[state]    (2)[above of=3]   {$2$};
\node[state]    (5)[right of=4]   {$5$};
\node[state]    (6)[right of=5]   {$6$};
\node[state]    (7)[right of=6]   {$7$};
\node[state]    (8)[right of=7]   {$8$};
\node[state]    (9)[right of=8]   {$9$};
\path
(1) edge[bend right]   node{$1$}    (7)
(2) edge[bend right]      node{$1$}     (1)
(3) edge[bend right]         node{$1$}     (1)
(4) edge         node{$1$}     (1)
(5) edge[bend right]    node{$1/3$}     (2)
(5) edge[bend right]   node{$1/3$}     (3)
(5) edge    node{$1/3$}     (4)
(6) edge[bend right]  node{$1$}     (5)
(7) edge[bend right]         node{$1-\eta$}     (6)
    edge[loop above]  node{$\eta$}     (5)
(8) edge[loop above]  node{$1$}     (8)
(9) edge[loop above]  node{$1$}     (9);
\end{tikzpicture}
  \caption{A game without a 0-equilibrium.}
  \label{noeq2figure}
\end{figure}

\begin{example} \label{noeq2} \rm We present another game with time-homogeneous Markov chain without a 0-equilibrium.
Consider the game in Figure~\ref{noeq2figure}. 
Let $\eta\in(0,1/6)$ and $q\in (0,1/3)$. 
Let $p=(0,0,0,0,0,q(1-\eta),q\eta,\frac{1-q}{2},\frac{1-q}{2})$ be the initial probability distribution. Notice that in this example there is no transient state.
\end{example}

\textsc{Claim 1:} In any 0-equilibrium, at period 1 player 1 chooses state 6 with probability 1.

\textsc{Proof of Claim 1.} If at period 1 player 1 looks at state 6, he guarantees $q(1-\eta)+\frac{1-q}{2}>1/2$ by looking at period 3 at state 8 or 9. 

If at period 1 player 1 looks at state 1, 2, 3, 4, 5 or 7, then player 2 can find the object with probability $q(1-\eta)$ at period 2 by looking at state 5 and with probability $\frac{1-q}{2}$ at period 4 by looking at state 8 or 9. As $q(1-\eta)+\frac{1-q}{2}>1/2$, player 1 cannot get more than $1/2$. 

If at period 1 player 1 looks at state 8 (respectively, at state 9), then player 2 can guarantee $\frac{1-q}{2}$ by looking at state 9 (respectively, at state 8) at period 2 and then $q(1-\eta)\cdot\frac{2}{3}$ by looking at state 1 at period 4. 
As $\frac{1-q}{2}+q(1-\eta)\cdot\frac{2}{3}=\frac{1}{2}+q(1/6-\eta)>1/2$ as $\eta<1/6$, player 1 cannot get more than $1/2$. 

So, there can be no 0-equilibrium in which at period 1 player 1 places a positive probability on a state different from state 6. \qed

\textsc{Claim 2:} In any 0-equilibrium, at period 2 player 2 chooses state 6 with probability 1.

\textsc{Proof of Claim 2.} From Claim 1, we know that in a 0-equilibrium, player 1 looks at state 6 at period 1 with probability 1. 
If he does so, he finds the object with probability $q(1-\eta)$ at period 1. 
Then, under the condition that the object is not found, the object was in state 7, 8 or 9 with probability 1 at period 1 and the updated probability distribution of the object is
$\left(0,0,0,0,0,0,\tfrac{\eta q}{1-q(1-\eta)},\tfrac{1-q}{2[1-q(1-\eta)]},\tfrac{1-q}{2[1-q(1-\eta)]}\right)$. 
Then, the object follows the transition matrix and the probability distribution of the object at period 2 is 
\begin{align*}
    &\left(0,0,0,0,0,
\frac{\eta(1-\eta) q}{1-q(1-\eta)},
\frac{\eta^2 q}{1-q(1-\eta)},
\frac{1-q}{2[1-q(1-\eta)]},
\frac{1-q}{2[1-q(1-\eta)]}\right)\\
=&\left(0,0,0,0,0,q'(1-\eta),q'\eta,\frac{1-q'}{2},\frac{1-q'}{2}\right),
\end{align*}
where $q'=\frac{\eta q}{1-q(1-\eta)}<\frac{q/6}{1-1/3(1-0)}=\frac{q}{4}<q<1/3$ as $0<\eta<1/6$ and $0<q<1/3$. 
Thus, at period 2 player 2 is facing a similar situation as player 1 at period 1. Claim 2 follows from Claim 1. \qed

\textsc{Claim 3:} This game has no 0-equilibrium.

\textsc{Proof of Claim 3.} Assume by way of contradiction that the game has a 0-equilibrium. From Claim 1, player 1 plays state 6 at period 1. From Claim 2, player 2 plays state 6 at period 2. By repeating the same reasoning as in Claim 2, in a 0-equilibrium, the active player looks at state 6 with probability 1 at each period. Under this strategy profile, the object is found with probability lower than $q<\tfrac{1-q}{2}$. Hence, it would be profitable for player 1 to deviate and look at state 6 at period 1. In conclusion, there is no 0-equilibrium. \qed

\subsection{Existence of $\veps$-equilibrium}

In this subsection we are interested in the existence of $\veps$-equilibrium, where $\veps>0$. We show that there is an $\veps$-equilibrium in pure strategies for every search game, and for each $\veps>0$. 
The proof relies on existence results for $\veps$-equilibria in games with Borel measurable payoff functions (see the proof of Mertens and Neyman in \cite{mertens1990repeated}) and with lower semi-continous payoff functions (see \cite{flesch2010perfect} and \cite{flesch2016subgame}). 

\begin{theorem}\label{vepsequilibrium} Each competitive search game admits an $\veps$-equilibrium in pure strategies, for all $\veps>0$. 
\end{theorem}

\textit{Proof.}
Consider the Model [2] of a competitive search game in Section~\ref{model}. Note that
\begin{enumerate}
    \item this is a multiplayer perfect-information game,
    \item from Proposition~\ref{lsc} it follows that the payoffs are bounded and lower semi-continuous.
\end{enumerate}
Thus by applying Theorem 2.3 of \cite{flesch2010perfect}, or Mertens and Neyman's result in \cite{mertens1990repeated}, to the Model [2], the game admits an $\veps$-equilibrium in pure strategies for every $\veps>0$. \qed

\textbf{Revisiting Example~\ref{noeq1} and ~\ref{noeq2}.} 
In view of Theorem \ref{vepsequilibrium}, the game in Example \ref{noeq1} has an $\veps$-equilibrium in pure strategies for every $\veps>0$. We now present an (subgame perfect) $\veps$-equilibrium in pure strategies of this game, for all $\veps>0$.

Let $\veps>0$. The idea of the $\veps$-equilibrium in pure strategies described here is to choose state 1 for a long time as long as the other player does the same, and then to choose the most likely between state 3 or state 4 in the remaining game. More formally, for each $n\in \N$, let $(\sigma^n,\tau^n)$ be the pure strategy profile defined as follows. For all $t\in\N$, for all history $h_t$ at period $t$, for all $n\in\N$, let $f_t^n:H_t\to S$ be defined by
\[
f^n_t(h_t)=\left\{ \begin{array}{lll} 
                         \mbox{state } 1 &  \mbox{    } & \mbox{ if } h_t\in \{1,2\}^{t-1} \mbox{ and } t<n, \\
                         \mbox{state } 3 &  \mbox{    } & \mbox{ if } h_t(t-1)= 4 \\ 
                         \mbox{      }   &  \mbox{ or } & \mbox{ if } h_t(t-1)\in \{1,2\} \mbox{ and } h_t(t-2)=4 \\
                         \mbox{      }   &  \mbox{ or } & \mbox{ if } h_t\in \{1,2\}^{t-1} \mbox{ and } t \geq n, \\
					     \mbox{state } 4 &  \mbox{    } & \mbox{ if } h_t(t-1)= 3 \\
					     \mbox{      }   &  \mbox{ or } & \mbox{ if } h_t(t-1)\in \{1,2\} \mbox{ and } h_t(t-2)=3.
					 \end{array}
			\right.
\]
Then, we define $\sigma^n_t(h_t)=f^n_t(h_t)$ for all $t\in \Nodd$, and $\tau^n_t(h_t)=f^n_t(h_t)$ for all $t\in \Neven$ and all history $h_t$ at time $t$.
The idea of $\sigma^n$ and $\tau^n$ is to look at state 1 until period $n$ (if the other player does the same) and from period $n$ onward (or before if the other player deviates) to look at the most likely state.
We argue that if $n\geq \tfrac{\ln{\eta q}-\ln{4\veps}}{\ln{2}-\ln{\eta}}$ then $(\sigma^n,\tau^n)$ is an $\veps$-equilibrium. For simplicity, we assume that $n$ is odd. 

It follows from the Claim 2 of the proof of Theorem \ref{noeqthm} that $\tau^n$ is a 0-best-response against $\sigma^n$. It is then sufficient to show that $\sigma^n$ is an $\veps$-best response against $\tau^n$ when $n$ is large enough.
From Claim 2 of the proof of Theorem \ref{noeqthm} it follows that a 0-best response against $\tau^n$ is to follow the strategy $\sigma^{n+1}$, which only differs from $\sigma^n$ at period $n$.
Under $(\sigma^n,\tau^n)$, player 1 finds the object at period 1 with probability $q$, player 2 finds the object at period 2 with probability $q.\left(\frac{\eta}{2}\right)$, player 1 finds the object at period 3 with probability $q.\left(\frac{\eta}{2}\right)^2$, and so on until period $n-1$ where player 2 finds the object with probability $q.\left(\frac{\eta}{2}\right)^{n-2}$. 
Then in the continuation game that starts at period $n$ it follows from the proof of Claim 2 in Theorem~\ref{noeqthm} that both players find the object with probability 1/2. 
So, player 1 finds the object before period $n$ with probability $q+\left(\frac{\eta}{2}\right)^2\cdot q+\left(\frac{\eta}{2}\right)^4\cdot q+\ldots+\left(\frac{\eta}{2}\right)^{n-3}\cdot q$, 
player 2 finds the object before period $n$ with probability $q.\left(\frac{\eta}{2}\right)+\ldots+q.\left(\frac{\eta}{2}\right)^{n-2}$ 
and each player finds the object from period $n$ with probability $ \frac{1}{2} \cdot \left[1-\left(q+\left(\frac{\eta}{2}\right)\cdot q+\left(\frac{\eta}{2}\right)^2\cdot q+...+\left(\frac{\eta}{2}\right)^{n-2}\cdot q\right)\right]$.
This implies that under $(\sigma_n,\tau_n)$ the expected payoff of player 1 is 
\[q+\left(\frac{\eta}{2}\right)^2\cdot q+\left(\frac{\eta}{2}\right)^4\cdot q+\ldots+\left(\frac{\eta}{2}\right)^{n-3}\cdot q+\left[1-\left(q+\left(\frac{\eta}{2}\right)\cdot q+\left(\frac{\eta}{2}\right)^2\cdot q+...+\left(\frac{\eta}{2}\right)^{n-2}\cdot q\right)\right] \cdot \frac{1}{2}\]
and under $(\sigma^{n+1},\tau^{n+1})$, the expected payoff of player 1 is
\[q+\left(\frac{\eta}{2}\right)^2\cdot q+\left(\frac{\eta}{2}\right)^4\cdot q+\ldots+\left(\frac{\eta}{2}\right)^{n-1}\cdot q+\left[1-\left(q+\left(\frac{\eta}{2}\right)\cdot q+\left(\frac{\eta}{2}\right)^2\cdot q+...+\left(\frac{\eta}{2}\right)^{n-1}\cdot q\right)\right] \cdot \frac{1}{2}.\]
Those two terms converge to the same limit  $\frac{q}{1-\left(\frac{\eta}{2}\right)^2}+\left[1-\frac{q}{1-\left(\frac{\eta}{2}\right)}\right]\cdot \frac{1}{2}$ which is the value of the game. 
Moreover, difference between these two expressions is \[\left|\left(\frac{\eta}{2}\right)^{n-1}\cdot q-\frac{1}{2}\cdot\left(\frac{\eta}{2}\right)^{n-1}\cdot q\right|=\frac{1}{2}\cdot\left(\frac{\eta}{2}\right)^{n-1}\cdot q.\] 
Hence, when $n\geq \tfrac{\ln{\left(\tfrac{\eta q}{4\veps}\right)}}{\ln{\left(\tfrac{2}{\eta}\right)}}$, the difference between those two expressions is smaller than $\veps$ so $(\sigma^n,\tau^n)$ is an $\veps$-equilibrium. \qed

With the same idea one can construct an $\veps$-equilibrium in Example \ref{noeq2} where both players choose state 6 until for a long time and then switch to state 8 or 9.

\subsection{Time-homogeneous Markov chains}

\label{timehomogeneous}

In this subsection, we consider time-homogeneous competitive search games. A game is time-homogeneous when the transition matrix $P_t$ at each period is the same. In this case, we will denote the transition matrix at each period by $P$. For all $r\in \N$ we denote by $P^r$, the matrix $P$ applied $r$ times.

Recall that a transition matrix $P$ is \textit{irreducible} if for each entry $(i,j)$, there exists $r\in\N$ such that the entry $(i,j)$ of $P^r$ is positive.
A transition matrix $P$ is \textit{periodic} of period $r\geq 2$ if for all $k\in \N$, $P^k(x,x)>0$ only if $k=r\cdot l$ for some $l\in \N$. If $P$ is not periodic, we say that $P$ is \textit{aperiodic}. 
A subset $S'\subseteq S$ is \textit{ergodic} if for $(i,j)\in S'\times (S\backslash S')$, $P(i,j)=0$ and the transition matrix $P$ restricted to the set $S'$ is irreducible. A state $i\in S$ is called \textit{absorbing} if $P(i,i)=1$. A state $i\in S$ is \textit{transient} if $\lim_{r\to \infty} P^r(i,i)=0$.

A probability distribution $\pi\in \Delta(S)$ over the set $S$ is called a  \textit{stationary distribution} for the transition matrix $P$ if $\pi P=\pi$.

It is known that (see \cite{levin2017markov}, Corollary 1.17  and Theorem 4.9) if the transition matrix $P$ is irreducible, then there exists a unique stationary distribution $\pi\in \Delta(S)$. If $P$ is also aperiodic, then there exist constants $\beta\in (0, 1)$ and $c > 0$ such that for all $t\in \N$,
\[ ||pP^t-\pi||_{TV}\leq c\cdot \beta^t,\]
where $||p-q||_{TV}=\underset{A\subset S}{\max}\sum_{s\in A} (p(s)-q(s))$ is the total variation distance over $\Delta(S)$.

\begin{theorem}\label{ergodic}
Consider a time-homogeneous competitive search game. Assume that the transition matrix $P$ is irreducible and aperiodic. Then, no matter the initial probability distribution $p$, every strategy profile finds the object with probability 1. Hence, the payoff functions are continuous in this game, and there exists a 0-equilibrium in pure strategies. 
\end{theorem}

\textit{Proof.} As mentioned, the transition matrix $P$ has a unique stationary distribution $\pi\in \Delta(S)$ and $\pi(s)>0$ for all $s\in S$. Moreover, there exist constants $c>0$ and $\beta\in (0,1)$ such that $|pP^t(s)-\pi(s)|\leq c\cdot\beta^t$ for all $t\in\N$, for all $s\in S$ and for all $p\in \Delta(S)$. Hence, there exists $t^*\in \N$ with the following property: for all $p\in \Delta(S)$, for all $s\in S$, for all $t\geq t^*$, we have $(pP^{t})(s)>\frac{\delta}{2}$, where $\delta=\min_{s\in S} \pi(s)$. Without loss of generality we can assume that $t^*\geq 2$.

Let $\alpha=\frac{\delta}{4(t^*-1)}$. The proof is divided into four steps.

\textsc{Step 1:} Let $(\sigma,\tau)$ be a pure strategy profile, and let $(s_t)_{t\in \N}$ denote the induced sequence of actions. We show that the object is found during the first $t^*$ periods with probability at least $\alpha$. 

\textsc{Proof:} For each $t\in\N$, let $p_{t}=(p_t(s))_{s\in S}\in \Delta(S)$ denote the probability distribution of the location of the object at period $t$, conditional on not being found through the history $(s_1,\ldots,s_{t-1})$.  

If there is a period $t\leq t^*$ such that $p_t(s_t)\geq \alpha$, then under $(\sigma,\tau)$, the object is found at period $t$ with probability at least $\alpha$, if it has not been found before. Hence, the claim of step 1 is true.

Therefore, it suffices to show that if at each period $t\leq t^*-1$ we have $p_t(s_t)< \alpha$, then $p_{t^*}(s_{t^*})\geq \alpha$. So assume that at each period $t\leq t^*-1$ we have $p_t(s_t)< \alpha$. The idea of the calculation below is that, since the object is found with low probabilities at the first $t^*-1$ periods, the probability distribution for the object at period $t^*$ on condition that it is not found during the first $t^*-1$ periods is almost the same as the unconditioned probability distribution. That is, $p_{t^*}$ is close to $pP^{t^*-1}$, which is in turn close to the stationary distribution $\pi$.

Note that, if the players do not condition on the past, the probability distribution of the location of the object at period $t^*$ is simply $pP^{t^*-1}$. We have
\begin{align*}
||p_{t^*}-pP^{t^*-1}||_{TV}&\leq ||p_{t^*}-p_{t^*-1}P||_{TV}+||p_{t^*-1}P-pP^{t^*-1}||_{TV}\\
&=||p_{t^*-1}^{\neg s_{t^*-1}} P-p_{t^*-1}P||_{TV}+||p_{t^*-1}P-pP^{t^*-1}||_{TV}\\
&\leq ||p_{t^*-1}^{\neg s_{t^*-1}}-p_{t^*-1}||_{TV}+||p_{t^*-1}-pP^{t^*-2}||_{TV}\\
&=p_{t^*-1}(s_{t^*-1})+||p_{t^*-1}-pP^{t^*-2}||_{TV}\\
&< \alpha+||p_{t^*-1}-pP^{t^*-2}||_{TV}\\
&< \alpha\cdot (t^*-1) +||p_1-pP^0||_{TV}\\
&= \alpha\cdot (t^*-1)\\
&=\frac{\delta}{4}.
\end{align*}
Here, in the first inequality we used the triangle inequality. In the first equality, we used that $p_{t^*}=p_{t^*-1}^{\neg s_{t^*-1}}P$, as $p_{t^*-1}^{\neg s_{t^*-1}}$ is the probability distribution of the location of the object at period $t^*-1$ conditional on the fact that the object has not been found before period $t^*-1$ and that it is not in state $s_{t^*-1}$ at period $t^*-1$ after the history $(s_1,\ldots,s_{t^*-2})$ and not being in state $s_{t^*-1}$ at period $t^*-1$. The second inequality is true as $||qP-q'P||_{TV}\leq ||q-q'||_{TV}$ for all $q,q'\in\Delta(S)$. 
The second equality follows from the above interpretation of $p_{t^*-1}^{\neg s_{t^*-1}}$ and of the total variation norm.
The third inequality is due to the assumption that at each period $t\leq t^*-1$ we have $p_t(s_t)< \alpha$. The fourth inequality then follows by induction. The last two equalities are due to $p_1=p$ and the choice of $\alpha$.

Therefore, \[p_{t^*}(s_{t^*})\,\geq\,  (pP^{t^*-1})(s_{t^*})-||p_{t^*}-pP^{t^*-1}||_{TV}\,\geq\,\frac{\delta}{2}-\frac{\delta}{4}\,=\,\frac{\delta}{4}\,\geq\, \alpha.\]
This completes the proof of Step 1.

\textsc{Step 2:} Consider any strategy profile $(\sigma,\tau)$. We show that the object is found during the first $t^*$ periods with probability at least $\alpha$.

\textsc{Proof:} On the finite horizon $t^*$, each strategy can be equivalently represented as a mixed strategy, i.e. a probability distribution on the finite set of pure strategies on horizon $t^*$ (see for example \cite{maschler2013game}). Hence, Step 2 follows from Step 1. 

\textsc{Step 3:} Consider any strategy profile $(\sigma,\tau)$. We show that the object is found with probability 1 under $(\sigma,\tau)$. By Proposition \ref{lsc}, this will imply that the payoff functions are continuous in this game. 

\textsc{Proof:} By Step 2, the object is found during the first $t^*$ periods with probability at least $\alpha$. Since $t^*$ and therefore $\alpha$ do not depend on the initial distribution of the object, if the object is not found in the first $t^*$ periods, then it will be found between periods $t^*+1$ and $2t^*$ with probability at least $\alpha$. By repeating this argument, the object is found with probability 1 under $(\sigma,\tau)$. 

\textsc{Step 4:} We show that there exists a 0-equilibrium in pure strategies.\footnote{By Step 3, the payoffs in the game are continuous. Since there is perfect information in the model representation [2], it follows from \cite{fudenberg1983subgame} en \cite{harris1985existence} that there even exists a subgame perfect 0-equilibrium in pure strategies.}

\textsc{Proof:} In view of Theorem \ref{vepsequilibrium}, for each $n\in \N$, there exists a $\frac{1}{n}$-equilibrium $(\sigma^n,\tau^n)$ in pure strategies. Since $\Sigma$ and $\mathcal{T}$ are compact and metrizable, by taking a subsequence if necessary, we can assume that the sequence $(\sigma^n,\tau^n)_{n\in\N}$ converges to a strategy profile $(\sigma,\tau)$ in pure strategies as $n\to\infty$ . 

For each $n\in \N$, we have $u_1(\sigma^n,\tau^n)\geq u_1(\sigma',\tau^n)-\frac{1}{n}$ and $u_2(\sigma^n,\tau^n)\geq u_2(\sigma^n,\tau')-\frac{1}{n}$ for all $\sigma'\in\Sigma$ and $\tau'\in\mathcal{T}$. Since by Step 3 the payoff functions $u_1$ and $u_2$ are continuous, by taking the limits as $n\to\infty$, we obtain $u_1(\sigma,\tau)\geq u_1(\sigma',\tau)$ and $u_2(\sigma,\tau)\geq u_2(\sigma,\tau')$ for all $\sigma'\in\Sigma$ and $\tau'\in\mathcal{T}$. Hence, $(\sigma,\tau)$ is a 0-equilibrium in pure strategies. \qed

\begin{remark}
\rm Consider a time-homogeneous search game. If this game does not satisfy the condition of Theorem~\ref{ergodic}, i.e. the transition matrix is not irreducible or not aperiodic, then the conclusion of Theorem~\ref{ergodic} is no longer true, and there is even an initial probability distribution of the object and a strategy profile under which the object is found with probability zero. Indeed, if the transition matrix is not irreducible or not aperiodic, we distinguish the following three (not exclusive) situations: (i) If there is a transient state, then consider an initial probability distribution which places probability zero on every transient state and a strategy profile which always chooses a transient state. (ii) If there is more than 1 ergodic class, then consider an initial probability distribution which places probability 1 on an ergodic class and a strategy profile which always chooses a state in another ergodic class. (iii) If there is a periodic ergodic class, then consider an initial probability distribution which places probability 1 on a state. Then due to periodicity, at each period there is a state where the object is with probability zero (see Exercise 1.6 of \cite{levin2017markov}). So consider a strategy profile which always chooses such a state. 
\end{remark}

\section{Payoff properties under $\veps$-equilibrium and existence of the value} \label{payoffproperties}

Competitive search games are not constant-sum games, and the payoff functions are not continuous as mentioned in Proposition~\ref{lsc}. 
We will first show that if a player chooses an $\veps$-best response against the strategy of the other player, the payoffs almost add up to 1. 
Thus, the game is essentially constant-sum, so the notion of value becomes natural. 
Then, we show the existence of the value of these games, to finally prove existence of $\veps$-optimal strategies for both players for all $\veps>0$ and relate optimal strategies and equilibria.

\begin{lemma}\label{bestresponse} Consider a strategy $\tau$ for player 2. Let $\veps>0$. If the strategy $\sigma$ of player 1 is an $\veps$-best response against $\tau$, then under $(\sigma,\tau)$ the object is found with probability at least $1-\veps\cdot|S|$. In other words, 
\[u_1(\sigma,\tau)+u_2(\sigma,\tau)\geq 1-\veps\cdot |S|. \]
A similar statement holds with exchanged roles of the players.
\end{lemma}

\textit{Proof.} Note that the sequence of events $([t<\Theta< +\infty])_{t\in\N}$ is decreasing and its limit is the empty set. Thus, $(\mathbb{P}_{(\sigma,\tau)}(t<\Theta< \infty))_t$ is decreasing and converges to 0 as $t$ goes to $\infty$, by $\sigma$-additivity of probability measures.

Suppose that player 1 plays $\sigma$ against $\tau$.
Then player 1 finds the object with probability $u_1(\sigma,\tau)=\mathbb{P}_{(\sigma,\tau)}(\Theta\in \Nodd)$. 
Assume now that player 1 follows $\sigma$ until a certain odd period $T\in \N$, and then deviates from $\sigma$ by choosing a state uniformly from period $T+2$ onward, and denote this strategy by $\sigma'$. Then player 1 finds the object at period $T+2$ with probability $(1-\mathbb{P}_{(\sigma,\tau)}(\Theta\leq T+1))/|S|$. 
Thus, $u_1(\sigma',\tau)\geq \mathbb{P}_{(\sigma,\tau)}(\Theta\in \Nodd, \ \Theta\leq T)+(1-\mathbb{P}_{(\sigma,\tau)}(\Theta\leq T+1))/|S|$.
As $\sigma$ is an $\veps$-best response against $\tau$, it holds that
\begin{align*}
    \mathbb{P}_{(\sigma,\tau)}(\Theta\in\Nodd)=u_1(\sigma,\tau)&\geq u_1(\sigma',\tau)-\veps=\mathbb{P}_{(\sigma',\tau)}(\Theta\in\Nodd)-\veps.
\intertext{So, since $\sigma$ and $\sigma'$ are identical for $\Theta\leq T+1$ this implies}
    \mathbb{P}_{(\sigma,\tau)}(\Theta\in\Nodd,\Theta \geq T+2)
    &\geq \mathbb{P}_{(\sigma',\tau)}(\Theta\in\Nodd,\Theta \geq T+2)-\veps \\
    &\geq (1-\mathbb{P}_{(\sigma,\tau)}(\Theta \leq T+1))/|S|-\veps \\
    &=\mathbb{P}_{(\sigma,\tau)}(\Theta \geq T+2))/|S|-\veps \\
    &\geq \mathbb{P}_{(\sigma,\tau)}(\Theta\in\Nodd,\Theta \geq T+2))/|S| +\mathbb{P}_{(\sigma,\tau)}(\Theta=\infty)/|S|-\veps.
\intertext{It follows that}
    \mathbb{P}_{(\sigma,\tau)}(\Theta=\infty)&\leq \mathbb{P}_{(\sigma,\tau)}(\Theta\in\Nodd,\Theta \geq T+2))\cdot (|S|-1)+\veps\cdot|S|\\
    &\leq \mathbb{P}_{(\sigma,\tau)}(T+2 \leq \Theta < \infty))\cdot (|S|-1)+\veps\cdot|S|.
\end{align*}
As $(\mathbb{P}_{(\sigma,\tau)}(T+2 \leq \Theta < \infty))_T$ converges to 0 when $T$ goes to $\infty$, then $\mathbb{P}_{(\sigma,\tau)}(\Theta=\infty)\leq \veps\cdot |S|$. Thus, $u_1(\sigma,\tau)+u_2(\sigma,\tau)=\mathbb{P}_{(\sigma,\tau)}(\Theta<\infty)=1-\mathbb{P}_{(\sigma,\tau)}(\Theta=\infty)\geq 1-\veps\cdot |S|$. \qed

We denote $v_1=\sup_{\sigma\in\Sigma} \inf_{\tau\in \mathcal{T}} u_1(\sigma,\tau)$ and $v_2=\sup_{\tau\in \mathcal{T}}\inf_{\sigma\in\Sigma} u_2(\sigma,\tau)$.

\begin{prop}\label{valueexist} The following equalities  hold:
\begin{align}
\label{supinf1}v_1&=\adjustlimits\inf_{\tau\in \mathcal{T}}\sup_{\sigma\in\Sigma} u_1(\sigma,\tau),\\
\label{supinf2}v_2&=\adjustlimits\inf_{\sigma\in\Sigma}\sup_{\tau\in \mathcal{T}} u_2(\sigma,\tau).\\
\label{v1v2}v_1&+v_2=1.
\end{align}
\end{prop}

\textit{Proof.}
First we prove equality (\ref{supinf1}). In this equality, player 1 is maximizing $u_1(\sigma,\tau)$ and player 2 is minimizing the same expression. Note that $(\sigma,\tau)\mapsto u_1(\sigma,\tau)$ is bounded. Moreover, by Proposition~\ref{lsc}, it is lower semi-continuous, and hence Borel measurable. Now, equality (\ref{supinf1}) follows from \cite{martin1975borel}, \cite{martin1998determinacy} or Maitra and Sudderth (1998).

Equality (\ref{supinf2}) follows similarly.

We now show that $v_1+v_2\leq 1$.
Let $\veps>0$ and let $(\sigma,\tau)$ be an $\veps$-equilibrium. We have:
\begin{align*}
    u_1(\sigma,\tau)\geq \sup_{\sigma'} u_1(\sigma',\tau)-\veps\geq \inf_{\tau'}\sup_{\sigma'} u_1(\sigma',\tau)-\veps=v_1-\veps.
\end{align*}
Similarly, $u_2(\sigma,\tau)\geq v_2-\veps$. Then,
\begin{align*}
    v_1+v_2\leq u_1(\sigma,\tau)+u_2(\sigma,\tau)+2\cdot\veps\leq 1+2\cdot\veps.
\end{align*}
As $\veps>0$ is arbitrary, we get $v_1+v_2\leq 1$.

We now show that $v_1+v_2\geq 1$. Let $\veps>0$ and let $(\sigma',\tau)$ be a strategy profile where $\sigma'$ is an $\veps$-best response against $\tau$. Then by Lemma~\ref{bestresponse} we have $u_1(\sigma,\tau)\geq 1-u_2(\sigma,\tau)-\veps\cdot|S|$. Denote $B^\tau_\veps\subseteq \Sigma$ the set of $\veps$-best responses of player 1 against $\tau$. We have
\begin{align*}
    v_1&=\adjustlimits\inf_{\tau\in \mathcal{T}}\sup_{\sigma\in\Sigma} u_1(\sigma,\tau)\\
    &\geq\adjustlimits\inf_{\tau\in \mathcal{T}}\sup_{\sigma \in B^\tau_\veps} u_1(\sigma,\tau)\\
    &\geq\adjustlimits\inf_{\tau\in \mathcal{T}}\sup_{\sigma \in B^\tau_\veps}\left[1-u_2(\sigma,\tau)-\veps\cdot|S|\right]\\
    &=1-\sup_{\tau\in \mathcal{T}}\inf_{\sigma\in B^\tau_\veps} u_2(\sigma,\tau)-\veps\cdot|S|\\
    &\geq 1-\sup_{\tau\in \mathcal{T}}\inf_{\sigma\in \Sigma} u_2(\sigma,\tau)-\veps\cdot|S|\\
    &=1-v_2-\veps\cdot|S|.
\end{align*}
As $\veps>0$ is arbitrary, we conclude that $v_1+v_2\geq 1$.
\qed

The last theorem of this section shows that all $\veps$-equilibria give almost the same payoffs, for small $\veps$.

\begin{theorem} \label{vepsnashvalue}
For each $\veps\geq 0$, for each $\veps$-equilibrium $(\sigma,\tau)$:
\begin{enumerate}[label={[\arabic*]}]
	\item the object is found with probability at least $1-\veps\cdot |S|$,
	\item $|u_1(\sigma,\tau)-v_1| \leq \veps$ and $|u_2(\sigma,\tau)-v_2| \leq \veps$,
\end{enumerate}
\end{theorem}
where $v_1$ and $v_2$ are characterised above Proposition~\ref{valueexist}.

\textit{Proof.}\ \\ \ 
[1] It is a direct consequence from Lemma~\ref{bestresponse}.

[2] Let $\veps\in (0,1)$. Let $(\sigma,\tau)$ be an $\veps$-equilibrium. As a consequence of Proposition~\ref{valueexist},
\[u_1(\sigma,\tau)\geq \sup_{\sigma' \in \Sigma} u_1(\sigma',\tau)-\veps \geq \underset{\sigma' \in \Sigma \ \tau' \in\mathcal{T}}{\sup \ \ \inf} \ u_1(\sigma',\tau')-\veps=v_1-\veps.\]
Similarly, $u_2(\sigma,\tau)\geq v_2-\veps$. Thus
\[u_1(\sigma,\tau)\leq 1-u_2(\sigma,\tau)\leq 1-(v_2-\veps)=v_1+\veps.\]
Similarly, $u_2(\sigma,\tau)\leq v_2+\veps$. Those inequalities give [2]. \qed

A competitive search game is not a constant sum game in a strict sense. However, Proposition \ref{valueexist} and Theorem \ref{vepsnashvalue} show that, in essence, it has the same properties as a game in which the payoffs add up to 1 and thus the players have opposite interest. This leads to the following definition.

\begin{definition} \label{vepsoptimaldef} Consider a competitive search game, and let $v_1$ and $v_2$ be as above Proposition~\ref{valueexist}.
\begin{enumerate}[label={[\arabic*]}]
	\item We call $v=v_1$ the value of the game. 
	\item For $\veps\geq 0$, we say that $\sigma\in \Sigma$ is an $\veps$-optimal strategy for player 1 if $u_1(\sigma,\tau)\geq v_1-\veps$ for every $\tau\in \mathcal{T}$. Similarly, we say that $\tau\in \mathcal{T}$ is an $\veps$-optimal strategy for player 2 if $u_2(\sigma,\tau)\geq v_2-\veps$ for every $\sigma\in \Sigma$.
\end{enumerate}

\end{definition}
For $\veps$-optimal strategies we obtain the following proposition.

\begin{prop}\label{vepsoptimalprop} Consider a competitive search game.
\begin{enumerate}[label={[\arabic*]}]
\item For all $\veps\geq 0$, if $(\sigma,\tau)$ is an $\veps$-equilibrium, then $\sigma$ and $\tau$ are $\veps$-optimal strategies.
\item For all $\veps\geq 0$, if $\sigma$ and $\tau$ are $\veps$-optimal strategies, then $(\sigma,\tau)$ is a $2\veps$-equilibrium.
\item A strategy profile $(\sigma,\tau)$ is a 0-equilibrium if and only if $\sigma$ and $\tau$ are 0-optimal strategies.
\item For all $\veps>0$, each player has a pure $\veps$-optimal strategy.
\end{enumerate} 
\end{prop}

\textit{Proof.}\ \\ \ 
[1] Let $(\sigma,\tau)$ be an $\veps$-equilibrium. Hence, $u_1(\sigma,\tau)\geq u_1(\sigma',\tau)-\veps$ for all $\sigma'\in \Sigma$. Then, $u_1(\sigma,\tau)\geq v_1-\veps$, which means that $\sigma$ is an $\veps$-optimal strategy for player 1. Similarly, $\tau$ is an $\veps$-optimal strategy for player 2.

[2] Assume now that $\sigma$ and $\tau$ are $\veps$-optimal strategies for player 1 and player 2. Let $\sigma'\in \Sigma$. Then, $u_2(\sigma',\tau)\geq v_2-\veps$. 
By Proposition~\ref{valueexist}, we get that 
\[u_1(\sigma',\tau)\, \leq \, 1-u_2(\sigma',\tau)\, \leq \, 1-(v_2-\veps)\, =\, v_1+\veps.\]
This implies that
$u_1(\sigma,\tau)\,\geq\, v_1-\veps\, \geq\, u_1(\sigma',\tau) -2\veps$.
Similarly, we obtain $u_2(\sigma,\tau)\geq u_2(\sigma,\tau')-2\veps$ for every $\tau'\in \mathcal{T}$
So, $(\sigma,\tau)$ is a $2\veps$-equilibrium.

[3] This is a direct consequence of [1] and [2].

[4] This is a consequence of [1] and Theorem~\ref{vepsequilibrium}.\qed

We end this section with a property of the value of time-homogeneous Markov chains. We show that if the initial probability distribution is exactly an invariant distribution of the transition matrix $P$, then player 1 has a weak advantage.

\begin{prop} \label{invariantdistribution}
Consider a time-homogenenous competitive search game. If $\pi$ is an invariant distribution of $P$, then $v(\pi)\geq 1/2$.
\end{prop}

\textit{Proof.} Assume first that there is a state $s\in S$ for which $\pi(s)=0$. Then $\pi^{\neg s}P=\pi P=\pi$. Since $\pi(s)=0$ we have  $v(\pi,s)=1-v(\pi)$. As $v(\pi)\geq v(\pi,s)$, we obtain $v(\pi)\geq 1-v(\pi)$. Hence, $v(\pi)\geq 1/2$. 

Assume there is no state $s\in S$ for which $p(s)=0$. Consider the game $G'$ that arises by adding a state $w$ to $G$. More precisely, $G'$ is the game with set of states $S'=S\cup\{w\}$, initial probability distribution $\pi'$ such that $\pi'(s)=\pi(s)$ for each state $s\in S$ and $\pi'(w)=0$, and transition matrix $P'$ that has the same transition probabilities between states in $S$ and makes $w$ absorbing.
Then, the object will never be in $w$ with probability 1. 
From Step 1 of the proof of [2] in Theorem~\ref{geometry}, the players may ignore state $w$ during the game. 
Then, $\pi'$ is an invariant distribution of $P'$, and hence by the first part we find $v(\pi)=v'(\pi')\geq 1/2$. \qed

\textbf{Remark.} We conjecture that if $P$ is irreducible and aperiodic, then $v_1(\pi)>1/2$. 
The value $v_1(p)$ can be smaller than 1/2 if $p$ is not the invariant distribution. Indeed, for example with three states, initial probability distribution $p=(1/3,1/3,1/3)$ and a transition matrix $P$ such that at the second period the object is in state 1 with probability 1.

\section{Additional results} \label{additionalresults}

\subsection{Subgame optimal strategies} 

An $\veps$-optimal strategy is a relevant solution concept, but it has the drawback that if the opponent makes a mistake, the continuation strategy does not have to be $\veps$-optimal.
Hence, in this subsection we examine \emph{subgame $\veps$-optimal} strategies. 

A strategy $\sigma$ for player 1 is called subgame $\veps$-optimal if, in each subgame, the continuation strategy of $\sigma$ is $\veps$-optimal. More precisely, for each history $h\in H^{\text{odd}}$ and strategy $\tau\in \mathcal{T}$ for player 2 
\[u_1(\sigma,\tau)(h) \geq v_1(h) -\veps.\]
The definition of a subgame $\veps$-optimal strategy for payer 2 is similar. Note that a subgame $\veps$-optimal strategy is also $\veps$-optimal.

\begin{example} \rm
In this example, we show that there are $\veps$-optimal strategies that are not subgame perfect $\veps$-optimal strategies. The set of states is $S=\{1,2\}$, the transition matrix $P$ is the identity over $S$ and the initial probability distribution is $p=(1,0)$. 

\begin{center}
\begin{tikzpicture}[->, >=stealth', auto, semithick, node distance=3cm]
\tikzstyle{every state}=[fill=white,draw=black,thick,text=black,scale=1]
\node[state]    (1)                  {$1$};
\node[state]    (2)  [right of=1]    {$2$};
\path
(1)   edge[loop above]    node{$1$}         (1)
(2)   edge[loop above]    node{$1$}         (2);
\end{tikzpicture}
\end{center}

The value of player 1 is $v_1=1$ and any optimal strategy of player 1 starts looking at state 1. 
Then, $v_2=0$ and all the strategies of player 2 are 0-optimal. In particular, it is optimal for player 2 to always choose state 2. Let $\tau$ denote this strategy.

Now suppose that player 1 makes a mistake and chooses state 2 at period 1. Then, the continuation strategy of $\tau$ from period 2 is not optimal. In fact, it would be the best for player 2 to choose state 1 at period 2 and win the game. \qed 
\end{example}

\begin{prop} \label{subgameperfectvepsequilibrium}
Consider a competitive search game. 
\begin{enumerate}
    \item For every $\veps>0$, each player has a pure strategy which is subgame $\veps$-optimal.
    \item Let $\veps\in(0,\frac{1}{|S|})$. If $\sigma$ is a subgame $\veps$-optimal strategy for player 1, then for every strategy $\tau$ of player 2, the object is found with probability 1 under the strategy profile $(\sigma,\tau)$. A similar statement holds for player 2.
\end{enumerate}
\end{prop}

\textit{Proof.}
[1] Let $\veps>0$. In \cite{flesch2010perfect} and \cite{flesch2016subgame} it is shown that there exists a subgame perfect $\veps$-equilibrium $(\sigma,\tau)$ in pure strategies. Now consider a subgame at a history $h$. Since the continuation strategies of $\sigma$ and $\tau$ at $h$ form an $\veps$-equilibrium, it follows similarly to Proposition~\ref{vepsoptimalprop} that the continuation strategy of $\sigma$ at $h$ is $\veps$-optimal in the subgame, and similarly the continuation strategy of $\tau$ at $h$ is $\veps$-optimal in the subgame.
Hence, $\sigma$ and $\tau$ are subgame $\veps$-optimal.

[2] Let $\veps\in(0,\frac{1}{|S|})$ and let $\sigma$ be a subgame $\veps$-optimal strategy. Consider a history $h$ at an odd period. The strategy for player 1 which looks at a state with the highest probability guarantees $1/|S|$ in the subgame at $h$. So, $v(h)\geq 1/|S|$. 

Now consider a strategy $\tau$ for player 2. Then, we have $u_1(\sigma,\tau)(h)\geq 1/|S|-\veps>0$. In particular, in the subgame at $h$, the object is found with probability at least $1/|S|-\veps>0$ under $(\sigma,\tau)$. Since this holds for every history $h$ at an odd period, by L\'{e}vy's zero-one law, the  object is found with probability 1 under $(\sigma,\tau)$. \qed

\subsection{Structure of the optimal actions} \label{structuralproperties}

In this subsection, we present some structural properties of the optimal actions. For all $s\in S$ and for all $p\in \Delta(S)$, we denote by $v_1(p)$ the value of the game with initial probability distribution $p$, and by $v_1(p,s)$ the expected payoff of player 1 if he chooses state $s$ at period 1 when the initial distribution is $p$, assuming that both players will play optimally afterwards. For each state $s\in S$, let $e^s\in \Delta(S)$ denote the probability distribution which allocates probability 1 to state $s$ and probability 0 on the other states. Thus, $v_1(e^s)=v_1(e^s,s)=1$ and for all $p\in\Delta(S)\backslash\{e^s\}$,
\[v_1(p,s) \, = \, p(s)+(1-p(s))\cdot(1-v_1(p^{\neg s}P)) \, = \, 1-(1-p(s))\cdot v_1(p^{\neg s} P),\]
where $p^{\neg s}$ is the probability distribution $p$ conditional to the fact that the object is not in state $s$. In other words, $p^{\neg s}(s)=0$ and $p^{\neg s}(j)=\frac{p(j)}{1-p(s)}$ for all $j\neq s$.
Note that $v_1(p)=\max_{s\in S} v_1(p,s)$.
We also denote for all $s\in S$ the set $A_s$ of the initial probability distributions for which it is optimal for player 1 to look at state $s$ at period 1. In other words, $A_s=\{p\in \Delta(S)\ | \ v_1(p,s)=v_1(p)\}$. Note that $\cup_{s\in S} A_s=\Delta(S)$.

\begin{theorem}\label{geometry} The optimality regions $A_s$ have the following properties.

[1] If the initial probability $p$ is sufficiently close to $e^s$, for some state $s$, then choosing state $s$ is the only optimal action. That is, the region $A_s\backslash \cup_{j\neq s} A_j$ is a neighborhood of $e^s$ in $\Delta(S)$.

[2] Looking at a state in which the object is with zero probability is never better than looking anywhere else. That is, for all states $s,s'\in S$, for all $p\in \Delta(S)$, if $p(s')=0$ then $v_1(p,s')\leq v_1(p,s)$. 

[3] For each subset $N\subseteq S$, the convex hull of the vertices $e^s$ with $s\in N$ is included in the set $\cup_{s\in N}A_s$.

[4] There is an initial distribution at which choosing any state is optimal. That is, $\cap_{s\in S} A_s\neq\emptyset$.

[5] For all $s\in S$, the region $A_s$ is star convex centered in $e^s$. That is, if $p\in A_s$ then the whole line segment between $p$ and $e^s$ is included in $A_s$.
\end{theorem} 

\textit{Proof.}

[1] The statement follows from the facts that each $v(p,s)$ is continuous (cf. Theorem~\ref{lip}) in $p$ and that $v(e^s,s)=1$ and $v(e^s,j)<1$ for all $j\neq s$. 

[2] Assume $p(s')=0$ for some state $s'\in S$. Let $s\in S$. Let $(\sigma,\tau)$ be a strategy profile such that $\sigma_1(\emptyset)=s$ and $\sigma$ and $\tau$ be Markov strategies : for each $t\in \Nodd$ (resp. $t\in \Neven$), $\sigma_t$ (resp. $\tau_t$) is constant over the set $H_t$. 
Let $\sigma'\in \Sigma$ be a Markov strategy of player 1 that starts looking at state $s'$. Let $p\in \Delta(S)$ and remark that $p=p(s)\cdot e^s +(1-p(s))\cdot p^{\neg s}$ for all $s\in S$, where $e^s\in \Delta(S)$ is the vector with $e^s(s)=1$ and $e^s(j)=0$ for all $j\neq s$. We have :
\begin{align*}
u_1(\sigma',\tau)(p)&=p(s)\cdot u_1(\sigma',\tau)(e^s)+(1-p(s)) \cdot u_1(\sigma',\tau)(p^{\neg s})\\
                      &\leq p(s)\cdot 1+(1-p(s)) \cdot u_1(\sigma',\tau)(p^{\neg s})\\
					  &=p(s) \cdot 1+(1-p(s)) \cdot u_1(\sigma,\tau)(p^{\neg s})\\
					  &=p(s) \cdot u_1(\sigma,\tau)(e^s)+(1-p(s)) \cdot u_1(\sigma,\tau)(p^{\neg s})\\
					  &=u_1(\sigma,\tau)(p).
\end{align*}
where the first equality comes from the linearity of the payoff function in respect of $p$ (see Section~\ref{linear}), the first inequality comes from the fact that the payoffs are bounded from above by 1, the second equality comes from the fact $[p^{\neg s}](s)=[p^{\neg s}](s')=0$ and that the game played will be the same as $\sigma,\sigma'$ and $\tau$ are not behavioral, the third equality comes from $u_1(\sigma,\tau)(e^s)=1$ as $\sigma_1(\emptyset)=s$, and the fourth equality comes from the linearity of the payoff in respect of $p$. Taking the supremum over $\sigma$ and the infimum over $\tau$ on both sides, we get $v_1(p,s')\leq v_1(p,s)$. \qed

[3] Let $p\in conv(\{e^s|s\in N\})$. Then $p(s)=0$ for all $s\notin N$. By [2], there is an optimal action $j\in N$, and hence $p\in \cup_{s\in N} A_s$.

[4] We will use the Knaster-Kuratowski-Mazurkiewicz (KKM) theorem\footnote{The KKM theorem states: 
Let $n\in \N$ be the cardinal of the set of states $S$, in other words $|S|=n$. Let $\Delta^n$ be the simplex in $\mathbb{R}^n$. A KKM covering is defined as a collection $C_1,\ldots, C_n$ of closed sets such that for any $N \subseteq \{1,\ldots, n\}$, the convex hull of the vertices corresponding to $N$ is covered by $\cup_{s\in N} C_s$. Then any KKM covering has a non-empty intersection, i.e.: $\cap_{s\in S} C_s \neq \emptyset$.}, see \cite{knaster1929beweis}. Note that by Theorem \ref{lip}, the function $p \mapsto v(p,s)$ is continuous for all $s\in S$. Thus, each region $A_s$ is closed. From this fact and from [3], we can apply the KKM theorem. We conclude from the KKM Theorem that $\cap_{s\in S} A_s \neq \emptyset$. 

[5] Let $s\in S$, let $p\in A_s$ and let $\lambda\in [0,1]$. We want to show that $\lambda e^s+(1-\lambda)p\in A_s$. Let $(\sigma,\tau)$ be a strategy profile. By equation (\ref{linearpayoff})
\begin{align*}
\sup_{\sigma} u_1(\sigma,\tau)(\lambda e^s+(1-\lambda)p)
&= \sup_\sigma \ \left[\lambda \cdot u_1(\sigma,\tau)(e^s)+(1-\lambda) \cdot u_1(\sigma,\tau)(p) \right]\\
&\leq \lambda \cdot  \left[\sup_\sigma\ u_1(\sigma,\tau)(e^s)\right] + (1-\lambda) \cdot \left[\sup_\sigma \ u_1(\sigma,\tau)(p)\right] \\
&=\lambda + (1-\lambda)\cdot \left[\sup_\sigma\ u_1(\sigma,\tau)(p)\right],
\end{align*}
where we used that $u_1(\sigma,\tau)(e^s)=1$ for any strategy $\sigma$ that looks at state $s$ at period 1. Hence
\[v(\lambda e^s+(1-\lambda)p)
=\inf_\tau\sup_\sigma u_1(\sigma,\tau)(\lambda e^s+(1-\lambda)p)
\leq \lambda + (1-\lambda)\cdot\left[\inf_\tau\sup_\sigma u_1(\sigma,\tau)(p)\right]
=\lambda + (1-\lambda)\cdot v(\sigma,\tau).\]
On the other hand, by theorem~\ref{linear}, $v(\lambda e^s+(1-\lambda)p,s)=\lambda +(1-\lambda)\cdot v(p,s)$. So, choosing $s$ when the initial probability distribution is $\lambda e^s+(1-\lambda)p$ is optimal. \qed

\begin{example}\label{geometrypicture}
\rm Consider the case in which the set of states is $S=\{1,2,3\}$. Let $Q=\begin{bmatrix}1&0&0\\0&1&0\\\frac{1}{2}&\frac{1}{2}&0\end{bmatrix}$. The sets $A_1$, $A_2$ and $A_3$ are represented in the time-homogeneous case where the transition matrix is the identity matrix in Figure~\ref{fig:test1}, and the matrix $Q$ in Figure ~\ref{fig:test2}. 

\begin{figure}[h]
  \centering
  \includegraphics[width=1\linewidth]{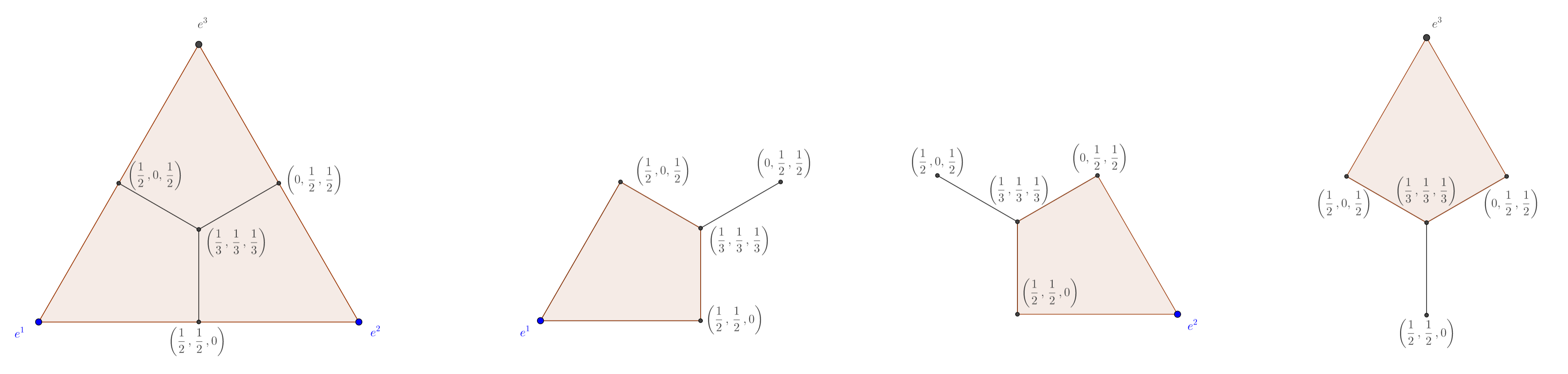}
  \captionof{figure}{$P=I_3$. From left to right, the sets $\Delta(S)$, $A_1$, $A_2$, $A_3$.}
  \label{fig:test1}
\end{figure}

\begin{figure}[h]
\centering
\includegraphics[width=0.5\linewidth]{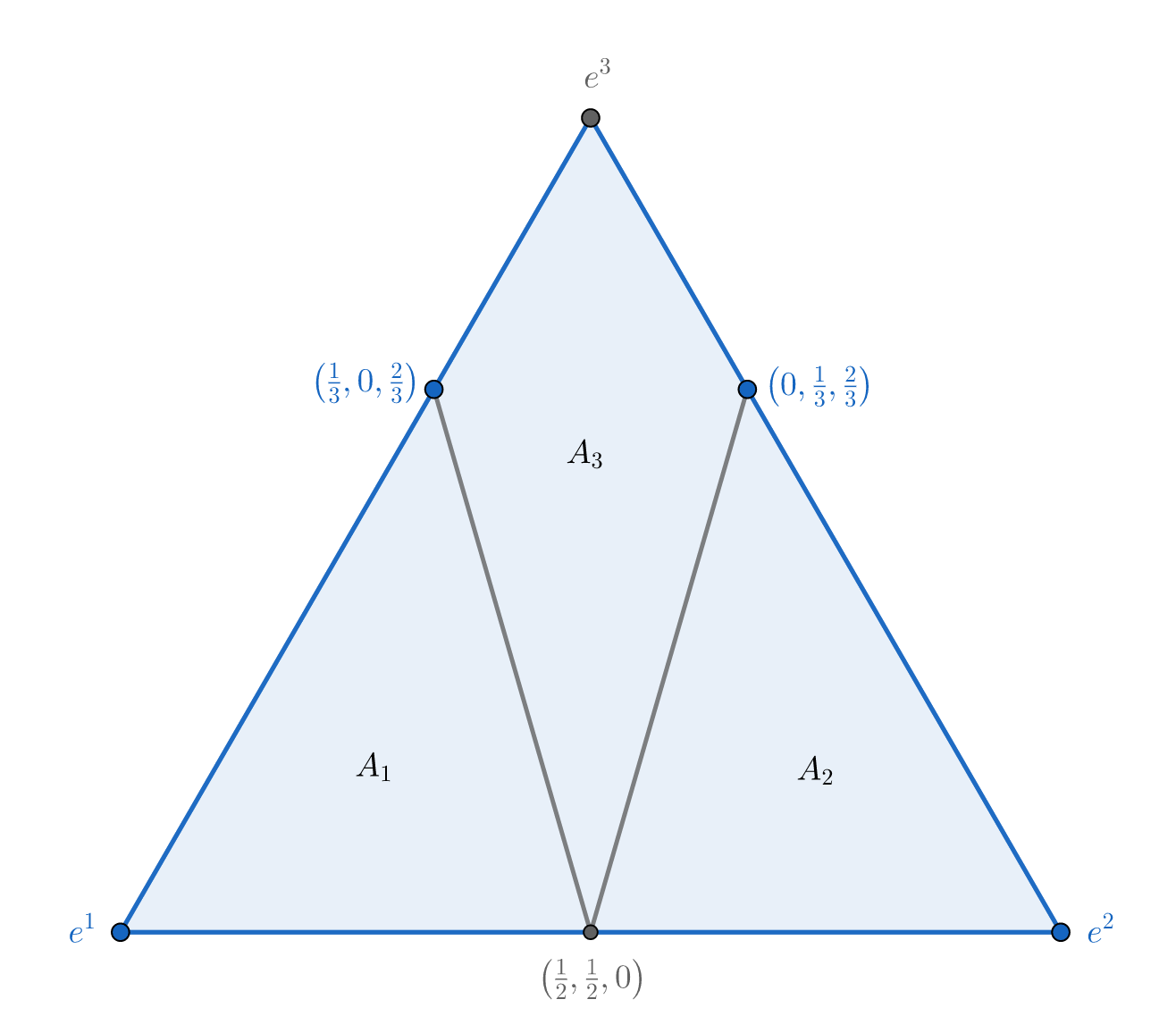}
\captionof{figure}{$P=Q$}
\label{fig:test2}
\end{figure}

\end{example}

Example~\ref{geometrypicture} illustrates the statements of Theorem~\ref{geometry}. It particular here are some remarks.

\begin{itemize}
\item It makes intuitive sense that if the object is in a certain state with probability close to 1, then it is optimal to look at this state. Geometrically, this means that for all states $s\in S$, the set $A_s$ contains a neighborhood of $e^s$ in $\Delta(S)$.
\item Looking at a state $s'$ such that $p(s')=0$ can still be (weakly) optimal. For example, in Figure \ref{fig:test1} with initial probability distribution $p=(1/2,1/2,0)$, looking at state 3 is just as good as looking at either state 1 or state 2.
\item Figure \ref{fig:test1} illustrates that the intersection of the regions $A_i$ can be more than a single point. 
\item Figure \ref{fig:test1} illustrates the sets $A_s$ are not always convex. However we conjecture that their relative interior is convex, in which case the closure of the relative interior of the sets $A_s$ are polytopes. 
\end{itemize}

\section{Variations} \label{strategies}

In this section we study two related versions of the search game: first where the horizon of the game is finite, and second through discounting when the players want to find the object as soon as possible. 
As we will see, the $\veps$-optimal strategies of the original model are robust, in the sense that they are $2\veps$-optimal if the horizon of the game is finite but sufficiently long, and they are also $2\veps$-optimal in the discounted version of the game, provided that the discount factor is close to 1. 
Similarly, each strategy that is optimal on a finite but sufficiently long horizon or for a high discount factor is also $\veps$-optimal in the original search game. 
In particular, as the optimal strategies over the finite horizon games can be calculated easily, we obtain $\veps$-optimal strategies in the original search game that are easy to calculate and to implement. 

\subsection{The finite horizon version of the search game}

Suppose that the game ends at a specific period $T\in\mathbb{N}$, if it has not ended before. For simplicity, we will focus on player 1. Let \[u_{1,T}(\sigma,\tau)=\mathbb{P}_{(\sigma,\tau)}(\Theta\in \Nodd, \ \Theta\leq T)\] denote the probability that player 1 finds the object within the $T$ first periods under $(\sigma,\tau)$. We assume that player 1 is maximizing $u_{1,T}$ whereas player 2 is minimizing $u_{1,T}$.
This is a zero-sum game which has value \[v_{1,T}:=\max_\sigma \min_\tau u_{1,T}(\sigma,\tau)= \min_\tau \max_\sigma u_{1,T}(\sigma,\tau).\]

Note that, with exchanged roles of the players, we could also define $v_{2,T}$.
However, since the game has finite horizon, it may have a positive probability under each strategy profile that the object is not found, so it will not always be true that $v_{1,T}+v_{2,T}=1$; in contrasts with Proposition~\ref{valueexist} for the infinite horizon. 

An advantage of the finite horizon compared to the infinite horizon is that the value in finite horizon can be computed explicitly via the following dynamic programming equations:
\begin{align*}
    v_{1,1}(p)&=v_{1,2}(p)=||p||_\infty,\\
    v_{1,T}(p)&=\max_{s_1}\min_{s_2}p(s_1)+(1-p(s_1)(1-[p^{\neg s_1}P_1](s_2))\cdot v_{1,T-2}([p^{\neg s_1}P_1]^{\neg s_2}P_2), \ T\geq 3 \mbox{ odd.}
\end{align*}

As we mentioned in the beginning of this section, the finite horizon search game is strongly related to the original search game.

\begin{definition} \rm
Let $\alpha\in(0,1)$. A transition matrix $P$ is $\alpha$-\textit{strongly mixed} if for all $(i,j)\in S\times S$, $P(i,j)\geq \alpha$.
\end{definition}

\begin{theorem}\label{theorem-fin-hor}  Consider a competitive search game.\\
{[1]} Let $\veps>0$. Let $\sigma^*\in\Sigma$ be an $\veps$-optimal strategy for player 1 in the original search game, and for all $T\in \N$, let $\sigma_T^*$ be a strategy for player 1 such that $u_{1,T}(\sigma_T^*,\tau)\geq v_{1,T}$ for each strategy $\tau$ of player 2. Then, there exists $\widetilde{T}\in \N$ such that for all $T\geq \widetilde{T}$, for all strategies $\tau\in \mathcal{T}$,
 \begin{equation}
  \label{inequality}
 u_{1,T}(\sigma^*,\tau)\geq v_1-2\veps \geq v_{1,T}-2\veps, \mbox{ and } u_{1,T}(\sigma_T^*,\tau)\geq v_1-\veps\geq v_{1,T}-\veps.
 \end{equation}
Consequently, $v_{1,T}$ converges to $v_1$ as $T$ goes to $\infty$.
  
{[2]} If there exists a real number $\alpha\in(0,1)$ such that for all $T\in \N$ the transition matrix $P_T$ at period $T$ is $\alpha$-strongly mixed, then for all $T\in\N$ 
\[v_1\geq v_{1,T}\geq v_1-(1-\alpha)^{T-1}.\]

{[3]} Analogous statements hold for player 2.
\end{theorem}
 
 \textit{Proof.}
 
\textsc{Proof of [1].} The second inequality in (\ref{inequality}) and the fourth inequality in (\ref{inequality}) are trivial. We now prove that for large $T$ the first inequality of (\ref{inequality}) holds. Assume by way of contradiction that for every $\Tilde{T}\in \N$, there exists $T\geq \Tilde{T}$ and there is a strategy $\tau_T$ such that $u_{1,T}(\sigma^*,\tau_T)< v_1-2\veps$. Since the set of strategies $\mathcal{T}$ for player 2 is compact, by taking a subsequence if necessary, we can assume that $\tau_T$ converges to some strategy $\tau$ as $T\to\infty$. Note that for every $T'\leq T$ we have
  \[u_{1,T'}(\sigma^*,\tau_T)\leq u_{1,T}(\sigma,\tau_T)< v_1-2\veps.\]
By taking the limit for $T\to \infty$, we find $u_{1,T'}(\sigma^*,\tau)\leq  v_1-2\veps$. Since this holds for all $T'$, when taking the limit for $T'\to \infty$, we obtain $u_1(\sigma^*,\tau)\leq v_1-2\veps<v_1-\veps$. This is a contradiction with the choice of $\sigma^*$. Thus, the inequality (\ref{inequality}) holds.

Now we prove that for large $T$ the third inequality of \ref{inequality} holds. Choose $\widetilde{T}$ so that the first inequality of \ref{inequality} holds for $\veps/2$. Then
\[ u_{1,T}(\sigma_T,\tau)\geq v_{1,T}=\max_{\sigma}\min_{\tau}u_{1,T}(\sigma,\tau)\geq \min_\tau u_{1,T}(\sigma,\tau)\geq v_1-\veps.\]
 
\textsc{Proof of [2].} The first inequality is trivial. Assume that there exists a real number $\alpha\in(0,1)$ such that for all $T\in \N$ the transition matrix $P_T$ at period $T$ is $\alpha$-strongly mixed. Let $T\in \N$.
 We use the following notations:
 \begin{itemize}
     \item $\sigma_T^*$ is an optimal strategy for player 1 in the zero-sum game with payoffs $(u_{1,T},-u_{1,T})$,
     \item $\sigma_T^-$ an optimal strategy for player 1 in the zero-sum game with payoffs $(-u_{2,T},u_{2,T})$,
     \item $\tau_T^*$ an optimal strategy for player 2 in the zero-sum game with payoffs $(-u_{2,T},u_{2,T})$,
     \item $\tau_T^-$ an optimal strategy for player 2 in the zero-sum game with payoffs $(u_{1,T},-u_{1,T})$.
 \end{itemize}
 Let $(\sigma,\tau)$ be a strategy profile. We have:
 \begin{align*}
     u_{1,T}(\sigma,\tau)+u_{2,T}(\sigma,\tau)
     &=u_{1,T-1}(\sigma,\tau)+u_{2,T-1}(\sigma,\tau)+\mathbb{P}_{(\sigma,\tau)}(\Theta=T)\\
     &\geq u_{1,T-1}(\sigma,\tau)+u_{2,T-1}(\sigma,\tau)+\left[1-u_{1,T-1}(\sigma,\tau)-u_{2,T-1}(\sigma,\tau) \right]\cdot \alpha\\
     &=\left(1-\alpha \right)\cdot \left[ u_{1,T-1}(\sigma,\tau)+u_{2,T-1}(\sigma,\tau)\right] +\alpha.
 \intertext{Then,}
     u_{1,T}(\sigma,\tau)+u_{2,T}(\sigma,\tau)-1
     &\geq (1-\alpha)\cdot\left[u_{1,T-1}(\sigma,\tau)+u_{2,T-1}(\sigma,\tau)-1\right],
  \intertext{which implies by induction}
     u_{1,T}(\sigma,\tau)+u_{2,T}(\sigma,\tau)-1
     &\geq(1-\alpha)^{T-1}\cdot\left[u_{1,1}(\sigma,\tau)+u_{2,1}(\sigma,\tau)-1\right]\\
     &=(1-\alpha)^{T-1}\cdot[p(\sigma(\emptyset))-1].
\intertext{Thus,}
     u_{1,T}(\sigma,\tau)+u_{2,T}(\sigma,\tau)
     &\geq 1-(1-\alpha)^{T-1}\cdot [1-p(\sigma(\emptyset))]\geq 1-(1-\alpha)^{T-1}.
\end{align*}
In particular,
\begin{align*}
     u_{1,T}(\sigma_T^*,\tau_{T}^-)+u_{2,T}(\sigma_{T}^-,\tau_T^*)
     &\geq u_{1,T}(\sigma_{T}^-,\tau_{T}^-)+u_{2,T}(\sigma_{T}^-,\tau_{T}^-)\\
     &\geq 1-(1-\alpha)^{T-1}\\
     &= v_1+v_2-(1-\alpha)^{T-1}
\end{align*}
As $u_{2,T}(\sigma_{T}^-,\tau_T^*)=v_{2,T}\leq v_2$, it implies 
\[ v_{1,T}\geq u_{1,T}(\sigma_T^*,\tau_{T}^-)\geq  v_1- (1-\alpha)^{T-1}.\]
\qed

\subsection{The discounted version of the search game}

Now we examine the discounted optimal strategies, once again with focus on player 1. 
For a discount factor $\beta\in(0,1)$ and strategy pair $(\sigma,\tau)$, let
\[u_{1,\beta}(\sigma,\tau)=\sum_{t\in \Nodd} \beta^{t-1}\mathbb{P}_{\sigma,\tau}(\Theta=t),\] 
which is the expected discounted time that player 1 finds the object, not counting the instances where the object is not found. 
We assume that player 1 is maximizing $u_{1,\beta}$ whereas player 2 is minimizing $u_{1,\beta}$. This is a zero-sum game. 
Let $v_{1,\beta}$ denote corresponding the value, and let $\sigma_\beta$ denote a pure optimal\footnote{In discounted games, one usually considers stationary strategies. In our model, the natural state space would be the set $\Delta(S)$ of possible probability distributions for the location of the object (often called the belief space, as the players only have a belief where the object could be). Since this space is infinite, and states are often only visited once, we omit the detailed discussion of stationarity.}
strategy of player 1. Note that the value and such a strategy $\sigma_\beta$ exist, because the discounted payoff is continuous (cf. for example Fudenberg and Levine (1983)). 
With exchanged roles of the players, we can also define $v_{2,\beta}$, and due to discounting we generally do not have $v_{1,\beta}+v_{2,\beta}=1$.

As we mentioned in the beginning of this section, the discounted search game is strongly related to the original search game.

\begin{theorem} \label{discount} Consider a competitive search game.\\ 
{[1]} Let $\veps>0$. Let $\sigma\in\Sigma$ be an $\veps$-optimal strategy for player 1, and for all $\beta\in(0,1)$, let $\sigma_\beta^*$ be a strategy for player 1 such that $u_{1,\beta}(\sigma_\beta^*,\tau)\geq v_{1,\beta}$ for each strategy $\tau$ of player 2. Then, there exists $\widetilde{\beta}\in (0,1)$ such that for all $\beta\in (\widetilde{\beta},1)$, for all strategies $\tau\in \mathcal{T}$,
\begin{equation*}
u_{1,\beta}(\sigma,\tau)\geq v_1-2\veps, \mbox{ and } u_{1}(\sigma_\beta^*,\tau)\geq v_1-\veps.
\end{equation*}
Consequently, $v_{1,\beta}\to v_1$ as $\beta\to 1$.

{[2]} Analogous statements hold for player 2.
 \end{theorem}
 
 \textit{Proof.}  

\textsc{Proof of [1].} For every $T\in\N$ let $\delta(T)\in(0,1)$ such that $(\delta(T))^{T-1}\geq 1-\tfrac{1}{T^2}$. Then, for every $\beta\in[\delta(T),1)$ and every strategy profile $(\sigma,\tau)$
\[u_{1,\beta}(\sigma,\tau)\geq \sum_{\substack{t=\Nodd\\  t\leq T}} \beta^{t-1}\cdot\mathbb{P}_{(\sigma,\tau)}(\Theta =t) \geq \sum_{\substack{t=\Nodd\\  t\leq T}} \left(1-\tfrac{1}{T^2}\right) \cdot\mathbb{P}_{(\sigma,\tau)}(\Theta =t) \geq u_{1,T}(\sigma,\tau)-\tfrac{1}{T}.\]
Hence, for all $\veps>0$, for all $T>\tfrac{1}{\veps}$, the statements of the theorem follow from Theorem \ref{theorem-fin-hor}. \qed

\section{Concluding remarks and future work} \label{conclusion}
We introduced an infinite horizon search game, in which two players compete to find an object that moves according to a time-varying Markov chain. We prove that these games always admit an $\veps$-equilibrium in pure strategies, for all error-terms $\veps>0$, but not necessarily a 0-equilibrium. 
We showed that the $\veps$-equilibrium payoffs converge to a singleton $(v, 1 - v)$ as $\veps$ vanishes, and therefore the game is essentially a zero-sum game with value $v$. 
We examined the analytical and structural properties of the solutions, and demonstrated that they are robust to having a finite but long horizon and respectively to having a sufficiently large discount factor. We devoted attention to the important special case when the Markov chain is time-homogeneous, where stronger results hold.

It would be interesting to generalize the results when the active player is chosen according to an arbitrary stochastic process. Also, one could introduce overlooking probabilities to the model. In that case, even if the active player chooses the state that currently contains the object, there is a positive probability that the player fails to find it. In the companion paper \cite{duvocelle2020search}, we examine the variation in which the active player is chosen randomly at each period.

\appendix

\section{Topological properties of search games}

We endow the strategy spaces $\Sigma=\prod_{h\in H^{\text{odd}}}\Delta(S)$ and $\mathcal{T}=\prod_{h\in H^{\text{even}}}\Delta(S)$ with the topology of pointwise convergence. This is identical with the product topology on $\Sigma$ and the product topology on $\mathcal{T}$. Under this topology, the spaces $\Sigma$ and $\mathcal{T}$ are compact, and as $H^{\text{odd}}$ and $H^{\text{even}}$ are countable, $\Sigma$ and $\mathcal{T}$ are also metrizable. 

\begin{definition} \rm
Let $X$ be a topological space. A function $f:X \to \R$ is called \textit{lower semi-continuous at} $x\in X$ if, for every sequence $x^n\to x$, we have $\liminf_{n\to\infty } f(x^n) \geq f(x)$. A function $f:X \to \R$ is called \textit{upper semi-continuous} at $x\in X$ if, for every sequence $x^n\to x$, we have $\limsup_{n\to\infty } f(x^n) \leq f(x)$. A function $f:X \to \R$ is called \textit{continuous} at $x\in X$ if it is lower semi-continuous at $x$ and upper semi-continuous at $x$.

A function $f:X \to \R$ is called \textit{lower semi-continuous} (resp. \textit{upper semi-continuous}, resp. \textit{continuous})  if $f$ is lower semi-continuous at all $x\in X$ (resp. upper semi-continuous at all $x\in X$, resp. continuous at all $x\in X$).
\end{definition}

\begin{prop} \label{lsc}
Take a player $i\in\{1,2\}$.
\begin{enumerate}[label={[\arabic*]}]
\item The payoff function $u_i:\Sigma\times \mathcal{T} \to \R$ is lower semi-continuous.
\item Assume that $(\sigma,\tau)$ is a strategy profile under which the object is found with probability 1. Then, $u_i$ is continuous at $(\sigma,\tau)$.
\end{enumerate}
\end{prop}

\textit{Proof.} \\
\textit{[1]} For each strategy profile $(\sigma,\tau)\in \Sigma\times \mathcal{T}$, for each period $n\in \N$, we denote by $u_i^n(\sigma,\tau)$ the probability that player $i$ finds the object during the first $n$ periods under the strategy profile $(\sigma,\tau)$. Note that $u_i^n(\sigma,\tau)$ is non-decreasing in $n$ and converges to $u_i(\sigma,\tau)$ as $n\to\infty$.

Let $(\sigma^k,\tau^k)_{k\in\N}$ be a sequence in $\Sigma \times \mathcal{T}$ converging to a strategy profile $(\sigma,\tau)$. We have for each $n\in\N$
\[ u_i^n(\sigma,\tau)=\lim_{k\to \infty} u_i^n(\sigma^k,\tau^k)=\liminf_{k\to \infty} u_i^n(\sigma^k,\tau^k)\leq \liminf_{k\to \infty} u_i(\sigma^k,\tau^k).\] 
Since $u_i^n(\sigma,\tau)$ converges to $u_i(\sigma,\tau)$ as $n\to\infty$, we obtain 
\[u_i(\sigma,\tau)\leq \liminf_{k\to \infty} u_i(\sigma^k,\tau^k),\]
which proves that $u_i$ is lower semi-continuous.\\
\textit{[2]} Assume that under the strategy profile $(\sigma,\tau)$ the object is found with probability 1. Thus, $u_1(\sigma,\tau)+u_2(\sigma,\tau)=1$. Due to part 1, we only need to show that $u_1$ and $u_2$ are upper semi-continuous at $(\sigma,\tau)$. We will prove it for $u_1$; the proof for $u_2$ is similar.

Let $(\sigma^k,\tau^k)_{k\in\N}$ be a sequence in $\Sigma \times \mathcal{T}$ converging to $(\sigma,\tau)$. Then
\[ \limsup_{k\to\infty} u_1(\sigma^k,\tau^k) = 1-\liminf_{k\to\infty} (1-u_1(\sigma^k,\tau^k)) \leq 1-\liminf_{k\to\infty} u_2(\sigma^k,\tau^k) \leq 1-u_2(\sigma,\tau) = u_1(\sigma,\tau), \]
where the first equality is a classic supinf equality applied to a limit, the first inequality comes from $u_1+u_2\leq 1$, the second inequality follows from part 1, and the second equality comes from the assumption we made on $(\sigma,\tau)$. Hence, $u_1$ is upper semi-continuous at $(\sigma,\tau)$, as desired. \qed

\section{Functional properties of the value function}

In this section we discuss some general functional properties of the value function $p\mapsto v(p)$. The first theorem is devoted to linear properties and the second theorem to Lipschitz-continuity. We remind that the function $p\mapsto v_1(p,s)$ was introduced at the beginning of the subsection~\ref{structuralproperties}.

\begin{theorem} \label{linear}
Let $(\sigma,\tau)$ be a strategy profile. Then the expected payoff functions are linear in the initial probability distribution of the object: for every $\lambda\in [0,1]$, for every $p,q\in \Delta(S)$, for every player $i=1,2$,
\begin{equation}\label{linearpayoff}
    u_i(\sigma,\tau)(\lambda p+(1-\lambda)q)= \lambda\cdot u_i(\sigma,\tau)(p)+(1-\lambda)\cdot u_i(\sigma,\tau)(q).
\end{equation}
Moreover, for every $s\in S$, the map $p\mapsto v(p,s)$ is linear over every line passing through $e^s$ (the initial probability distribution having probability 1 on state $s$): for every $p\in \Delta(S)$, for every $\lambda\in (0,1)$
\[v(\lambda e^s+(1-\lambda)p,s)=\lambda+(1-\lambda)\cdot v(p,s).\]
\end{theorem}

\textit{Proof.} First we prove equality (\ref{linearpayoff}). The probability distribution $\lambda\cdot p+(1-\lambda)\cdot q$ can be interpreted as follows: with probability $\lambda$ the initial probability distribution is $p$ and induces the expected payoff $u_i(\sigma,\tau)(p)$ for player $i$, and with probability $(1-\lambda)$ the probability distribution is $q$ and induces the expected payoff $u_i(\sigma,\tau,q)$ for player $i$. Hence, the equality (\ref{linearpayoff}) holds.

Now we prove the second part of the theorem.
Let $p\in \Delta(S)$, $p\neq e^s$, and let $\lambda \in (0,1)$, and denote $p^{\neg s}$ the linear projection of $x$ from $e^s$ to the face $\{y\in\Delta(S)| y_s=0\}$. Then 
\[(\lambda e^s+(1-\lambda)p)^{\neg s}=p^{\neg s}.\]
Indeed, $(\lambda e^s+(1-\lambda)p)^{\neg s}(s)=0=[p^{\neg s}](s)$ and for all $j \neq s$: 
\begin{align*}
(\lambda e^s+(1-\lambda)p)^{\neg s}(j)
&=\frac{(\lambda e^s+(1-\lambda)p)(j)}{1-(\lambda e^s+(1-\lambda)p)(s)}
=\frac{(1-\lambda)\cdot p(j)}{1-(\lambda +(1-\lambda)\cdot p(s))}\\
&=\frac{(1-\lambda)\cdot p(j)}{(1-\lambda)\cdot(1-p(s))}=\frac{p(j)}{(1-p(s))}=\frac{p(j)}{1-p(s)}=[p^{\neg s}](j).
\end{align*} 
Hence, by using $(\lambda e^s+(1-\lambda)p)(s)=\lambda+(1-\lambda)\cdot p(s)$ we have
\begin{align*}
&\ v(\lambda e^s+(1-\lambda)p,s)\\
&=(\lambda e^s+(1-\lambda)p)(s)+(1-(\lambda e^s+(1-\lambda)p)(s))\cdot(1-v((\lambda e^s+(1-\lambda)p)^{\neg s}P))\\
&=(\lambda e^s+(1-\lambda)p)(s)+(1-(\lambda e^s+(1-\lambda)p)(s))\cdot(1-v(p^{\neg s}P))\\
&=\lambda+(1-\lambda)(p(s)+(1-p(s))\cdot (1-v(p^{\neg s}P)))\\
&=\lambda+(1-\lambda)\cdot v(p,s),
\end{align*} 
which completes the proof. \qed

\textbf{Remark.} For each line passing through $e^s$, the linearity of the function $p\mapsto v(p,s)$ relies on the fact that if by choosing state $s$ player 1 does not find the object, then the conditional distribution of the location of the object, $p^{\neg s}$, stays on the same line. For lines not passing through $e^s$, this is no longer true, and the function $p\mapsto v(p,s)$ is generally non-linear. 
For example when $P=I_4$, $p=(1/3,1/3,1/3,0)$ and $p'=(1/3,1/3,0,1/3)$. In that case, $v(p,1)=2/3$ and $v(p',1)=2/3$, but $v_1(p/2+p'/2,1)=1/2$. 

Before introducing the next theorem, we recall the definition of the total variation distance: for $p,q\in \Delta(S)$, the total variation distance between $p$ and $q$ is the non-negative number
\[||p-q||_{TV}=\max_{S'\subset S}\sum_{s\in S'} [p(s)-q(s)].\]

\begin{theorem}\label{lip}
Let $p,q\in \Delta(S)$. Let $T\in \N$ and let $(\sigma,\tau)$ be a strategy profile. Then, the functions $p\mapsto u_{1,T}(\sigma,\tau)(p)$, $p\mapsto u_{1}(\sigma,\tau)(p)$, $p\mapsto v_{1,T}(p)$, $p\mapsto v_{1}(p,s)$ and $p\mapsto v_{1}(p)$ are $1$-Lipschitz continuous with respect to the total variation distance.
\end{theorem} 

\textbf{Proof.} By Theorem~\ref{linear}, we have \[u_1(\sigma,\tau)(p)=\sum_{s\in S}p(s)\cdot u_1(\sigma,\tau)(e^s),\] \[u_1(\sigma,\tau)(q)=\sum_{s\in S}q(s)\cdot u_1(\sigma,\tau)(e^s).\] Then,
\[u_1(\sigma,\tau)(p)-u_1(\sigma,\tau)(q)=\sum_{s\in S}[p(s)-q(s)]\cdot u_1(\sigma,\tau)(e^s)\leq \sum_{\substack{s\in S,\\ p(s)>q(s)}}[p(s)-q(s)]=||p-q||_{TV},\]
and similarly
\[u_1(\sigma,\tau)(q)-u_1(\sigma,\tau)(p)\leq||p-q||_{TV}.\]
Hence, $p\mapsto u_1(\sigma,\tau)(p)$ is $1$-Lipschitz-continuous.

Taking the infimum over $\tau$ and the supremum over $\sigma$ on both sides of the inequality $u_1(\sigma,\tau)(p)\leq u_1(\sigma,\tau)(q)+||p-q||_{TV}$ gives $v_1(p)\leq ||p-q||_{TV}+v_1(q)$, which can be written $v_1(p)-v_1(q)\leq ||p-q||_{TV}$. Similarly, $v_1(q)-v_1(p)\leq ||p-q||_{TV}$. Hence, $p\mapsto v_1(p)$ is $1$-Lipschitz-continuous too.

The proof for $p\mapsto u_{1,T}(\sigma,\tau)(p)$ and $p\mapsto v_{1,T}(\sigma,\tau)$ are similar. The proof for $p\mapsto v_{1}(p,s)$ is also similar, but the supremum in $\sigma$ has to be taken over the strategies that look at state $s$ at period 1. \qed

\bibliography{Acompetitivesearchgamewithamovingtarget.bib}{}

\begin{thebibliography}{10}

\bibitem{alpern2006theory}
Steve Alpern and Shmuel Gal.
\newblock {\em The theory of search games and rendezvous}, volume~55.
\newblock Springer Science \& Business Media, 2006.

\bibitem{benkoski1991survey}
Stanley~J Benkoski, Michael~G Monticino, and James~R Weisinger.
\newblock A survey of the search theory literature.
\newblock {\em Naval Research Logistics (NRL)}, 38(4):469--494, 1991.

\bibitem{brown1980optimal}
Scott~Shorey Brown.
\newblock Optimal search for a moving target in discrete time and space.
\newblock {\em Operations research}, 28(6):1275--1289, 1980.

\bibitem{dobbie1974two}
James~M Dobbie.
\newblock A two-cell model of search for a moving target.
\newblock {\em Operations Research}, 22(1):79--92, 1974.

\bibitem{duvocelle2020search}
Benoit Duvocelle, J\'anos Flesch, Hui~Min Shi, and Dries Vermeulen.
\newblock Search for a moving target in a competitive environment.
\newblock {\em arXiv preprint arXiv:2008.09653}, 2020.

\bibitem{flesch2009optimal}
J{\'a}nos Flesch, Emin Karag{\"o}zoǧlu, and Andr{\'e}s Perea.
\newblock Optimal search for a moving target with the option to wait.
\newblock {\em Naval Research Logistics (NRL)}, 56(6):526--539, 2009.

\bibitem{flesch2010perfect}
J{\'a}nos Flesch, Jeroen Kuipers, Ayala Mashiah-Yaakovi, Gijs Schoenmakers,
  Eilon Solan, and Koos Vrieze.
\newblock Perfect-information games with lower-semicontinuous payoffs.
\newblock {\em Mathematics of Operations Research}, 35(4):742--755, 2010.

\bibitem{flesch2016subgame}
J{\'a}nos Flesch and Arkadi Predtetchinski.
\newblock Subgame-perfect epsilon-equilibria in perfect information games with
  common preferences at the limit.
\newblock {\em Mathematics of Operations Research}, 41(4):1208--1221, 2016.

\bibitem{fudenberg1983subgame}
Drew Fudenberg and David Levine.
\newblock Subgame-perfect equilibria of finite-and infinite-horizon games.
\newblock {\em Journal of Economic Theory}, 31(2):251--268, 1983.

\bibitem{gal1979search}
Shmuel Gal.
\newblock Search games with mobile and immobile hider.
\newblock {\em SIAM Journal on Control and Optimization}, 17(1):99--122, 1979.

\bibitem{gal2010search}
Shmuel Gal.
\newblock Search games.
\newblock {\em Wiley Encyclopedia of Operations Research and Management
  Science}, 2010.

\bibitem{gal2013search}
Shmuel Gal.
\newblock Search games: a review.
\newblock In {\em Search Theory}, pages 3--15. Springer, 2013.

\bibitem{garnaev2012search}
Andrey Garnaev.
\newblock {\em Search games and other applications of game theory}, volume 485.
\newblock Springer Science \& Business Media, 2012.

\bibitem{garrec2020search}
Tristan Garrec and Marco Scarsini.
\newblock Search for an immobile hider on a stochastic network.
\newblock {\em European Journal of Operational Research}, 283(2):783--794,
  2020.

\bibitem{harris1985existence}
Christopher Harris.
\newblock Existence and characterization of perfect equilibrium in games of
  perfect information.
\newblock {\em Econometrica: Journal of the Econometric Society}, pages
  613--628, 1985.

\bibitem{hohzaki2016search}
Ryusuke Hohzaki.
\newblock Search games: Literature and survey.
\newblock {\em Journal of the Operations Research Society of Japan},
  59(1):1--34, 2016.

\bibitem{jordan1997optimal}
Benjamin~Paul Jordan.
\newblock {\em On optimal search for a moving target}.
\newblock PhD thesis, Durham University, 1997.

\bibitem{kan1974counterexample}
YC~Kan.
\newblock A counterexample for an optimal search-and-stop model.
\newblock {\em Operations Research}, 22(4):889--892, 1974.

\bibitem{knaster1929beweis}
Bronis{\l}aw Knaster, Casimir Kuratowski, and Stefan Mazurkiewicz.
\newblock Ein beweis des fixpunktsatzes f{\"u}r $n$-dimensionale simplexe.
\newblock {\em Fundamenta Mathematicae}, 14(1):132--137, 1929.

\bibitem{levin2017markov}
David~A Levin and Yuval Peres.
\newblock {\em Markov chains and mixing times}.
\newblock American Mathematical Society, 2017.

\bibitem{macphee1995optimal}
IM~MacPhee and BP~Jordan.
\newblock Optimal search for a moving target.
\newblock {\em Probability in the Engineering and Informational Sciences},
  9(2):159--182, 1995.

\bibitem{martin1975borel}
Donald~A Martin.
\newblock Borel determinacy.
\newblock {\em Annals of Mathematics}, 102(2):363--371, 1975.

\bibitem{martin1998determinacy}
Donald~A Martin.
\newblock The determinacy of blackwell games.
\newblock {\em The Journal of Symbolic Logic}, 63(4):1565--1581, 1998.

\bibitem{maschler2013game}
M~Maschler, Eilon Solan, and Shmuel Zamir.
\newblock Game theory.
\newblock {\em Cambridge University Press, Cambridge}, 2013.

\bibitem{mertens1990repeated}
Jean-Fran{\c{c}}ois Mertens.
\newblock Repeated games.
\newblock In {\em Game Theory and Applications}, pages 77--130. Elsevier, 1990.

\bibitem{nakai1973model}
TERUHISA Nakai.
\newblock Model of search for a target moving among three boxes: Some special
  cases.
\newblock {\em Journal of Operations Research Society of Japan}, 16:151--162,
  1973.

\bibitem{nakai1986search}
Teruhisa Nakai.
\newblock A search game with one object and two searchers.
\newblock {\em Journal of applied probability}, 23(3):696--707, 1986.

\bibitem{pollock1970simple}
Stephen~M Pollock.
\newblock A simple model of search for a moving target.
\newblock {\em Operations Research}, 18(5):883--903, 1970.

\bibitem{schweitzer1971threshold}
Paul~J Schweitzer.
\newblock Threshold probabilities when searching for a moving target.
\newblock {\em Operations Research}, 19(3):707--709, 1971.

\bibitem{stone1976theory}
Lawrence~D Stone.
\newblock {\em Theory of optimal search}, volume 118.
\newblock Elsevier, 1976.

\bibitem{stone2016optimal}
Lawrence~D Stone, Johannes~O Royset, and Alan~R Washburn.
\newblock Optimal search for moving targets (international series in operations
  research \& management science 237).
\newblock {\em Cham, Switzerland: Springer}, 2016.

\bibitem{washburn1983search}
Alan~R Washburn.
\newblock Search for a moving target: The fab algorithm.
\newblock {\em Operations research}, 31(4):739--751, 1983.

\end{thebibliography}
\bibliographystyle{plain}
\end{document}